\documentclass[preprint,aip,pop]{revtex4-1} 
\usepackage [english] {babel}
\usepackage [dvips] {graphicx}
\usepackage {amsmath}
\usepackage {amsfonts}
\usepackage {amssymb}
\usepackage {amsthm}
\usepackage{bm}
\usepackage{hyperref}
\usepackage{datetime}

\newcommand{\unit}[1]{\ensuremath{\, \mathrm{#1}}}

\newcommand{\bhat}{\bm{\hat{b}}}

\newcommand{\zhat}{\bm{\hat{z}}}

\newcommand{\pfrac}[2]{\frac{\partial#1}{\partial#2}}

\newcommand{\curl}[1]{\nabla \times #1}
\newcommand{\np}{\nabla_{\perp}}

\newcommand{\ExB}{$\bm{E}\times\bm{B}$}

\newcommand{\GA}[1]{\langle #1	 \rangle}

\newcommand{\reff}[1]{(\ref{#1})}

\begin{document}
\preprint{}

\title{Collisional transport across the magnetic field in drift-fluid models}

\author{J. Madsen}
\email{jmad@fysik.dtu.dk}
\author{V.~Naulin}
\author{A.~H. Nielsen}
\author{J.~Juul Rasmussen} 

\affiliation{Department of Physics, Technical University of Denmark,
DK-2800 Kgs. Lyngby,
Denmark}

\date{\today}
\currenttime

\begin{abstract}  
Drift ordered fluid models are widely applied in studies of low-frequency turbulence in the edge and scrape-off layer regions of magnetically confined plasmas. Here, we show how collisional transport across the magnetic field is self-consistently incorporated into drift-fluid models without altering the drift-fluid energy integral. We demonstrate that the inclusion of collisional transport in drift-fluid models gives rise to diffusion of particle density, momentum and pressures in drift-fluid turbulence models and thereby obviate the customary use of artificial diffusion in turbulence simulations. We further derive a computationally efficient, two-dimensional model which can be time integrated for several turbulence de-correlation times using only limited computational resources. The model describes interchange turbulence in a two-dimensional plane perpendicular to the magnetic field located at the outboard midplane of a tokamak. The model domain has two regions modeling open and closed field lines. The model employs a computational expedient model for collisional transport. Numerical simulations show good agreement between the full and the simplified model for collisional transport.
\end{abstract}
\maketitle

\section{Introduction}
There are basically two approaches to modeling the edge and scrape-off layer (SOL) regions in magnetically confined plasmas: mean field transport models and turbulence models. In the first approach\cite{schneider_solps_2006,radford1996EDGE2d}, mean field fluid equations are solved in a realistic geometry typically including the divertor and x-point regions. These models have very detailed descriptions of collisions, neutral particles, impurities, and material surfaces, whereas turbulent transport is not calculated self-consistently but modeled by effective turbulent diffusion terms. Turbulence models\cite{Garcia_POP_062309,Scott1997PPCF,RogersPRL1998,Ricci_PPCF_2012,Tamain2010,Thyagaraja2004475} take a different approach. Here, a detailed description of the turbulence is the main goal, whereas the level of detail of collisions, neutral particles, impurities and material surfaces, if included in the models at all, is crude compared with the highly detailed descriptions in mean field transport models. Furthermore, in order to reduce the computational requirements the magnetic field geometry is often simplified. Both approaches have provided useful results in their respective regimes of validity. Nevertheless, several open questions in contemporary fusion plasma research require that these complementary approaches are merged. Examples are: The formation of a elevated density shoulder\cite{MCCORMICK_JNM1992,GarciaJNM2007575} as the particle density approaches the Greenwald density limit and the associated changes of filament properties\cite{carralero_NF_2014}, the high particle density front\cite{Potzel2015541} on the high field side of single null diverted plasmas, and eventually the possible influence of turbulent transport on the transition to the detached\cite{stangeby2000plasma} divertor regime. 

Here, we present a step towards more detailed edge and SOL turbulence models by showing how classical collisional transport across the magnetic field is included in drift-fluid models\cite{Hinton_Horton_1971, Zeiler} based on the Braginskii fluid closure\cite{Braginskii}. Compared with neoclassical transport, classical collisional transport across the magnetic field is small in most regions of a magnetically confined fusion plasma. In turbulence dominated systems collisional transport is important because of its diffusive nature which inevitably determines the smallest possible scale length of the turbulence below which the turbulent energy is dissipated. It is well-known that turbulence behaves remarkably different when collisional dissipation is completely neglected compared with the situation where a very small but finite dissipation is present. The results of neoclassical transport theory are concerned with the steady-state and does not take turbulence into account nor resolve the much faster turbulent time-scale. In the Pfirsch-Schl\" uter regime, neoclassical transport is a result of collisional diffusive transport along the magnetic field in combination with guiding-center drifts due to magnetic field inhomogeneity. On the turbulent time-scale, it is therefore not evident that neoclassical effects through parallel collisional transport is the dominant regulator of turbulence in particular when the spatial scale of the turbulence is small as in e.g., drift wave turbulence. A consistent model for perpendicular collisional transport may therefore be of importance for the interplay between collisions and turbulence and certainly provide a valuable  validation tool for models which do not take classical perpendicular collisional transport into account. Moreover, classical perpendicular transport is believed to play a role for the perpendicular convection of filaments in the SOL region. A filament at the outboard midplane typically has a size which is on the order of $10-100$ ion gyroradii, but when then filament approaches the x-point region the perpendicular scale length can easily be smaller than the ion gyroradius due the strong magnetic shear near the x-point\cite{Farina1993}. In that case the perpendicular conductivity and viscosity are important mechanisms for closing the filament current loop\cite{Cohen2007} and thereby modify the filament convection and possibly decouple perturbations above and below the x-point. Finally, classical perpendicular transport are important in other fully ionized plasmas without neoclassical transport, e.g. cylindrical plasmas.

In drift-fluid models, the components of the momentum density equations perpendicular to the magnetic field are solved iteratively essentially under the assumption that the dynamics is evolving much slower than the  ion gyration time scale. In this limit, every force density, including resistivity and collisional viscosity, gives rise to a perpendicular drift\cite{HintonHazRevModPhys1976,Hinton1984}. Most of the collisional drifts are included in transport models\cite{schneider_solps_2006,radford1996EDGE2d} but not in a way which provides energy conservation. In turbulence models, on the other hand, the collisional drifts are usually neglected or do only include selected collisional effects\cite{DrakePOF1984}. In a model including ion temperature dynamics, we show how these collisional terms in combination with heat fluxes, heat exchange terms, and viscous heating terms in the electron and ion pressure equations are included in drift-fluid turbulence models in a consistent way. Consistency here mainly refers to energy conservation. In this paper we show that collisional transport across the magnetic field can be included in drift-fluid models without altering the energy theorem\cite{Zeiler,pop2003BDS}. For example, we show that the collisional viscous damping of turbulence and mean flows gives rise to a conservative energy transfer between the kinetic energy and the ion thermal energy. In other words, collisions can be added to drift-fluid models without introducing energy sinks or sources. The inclusion of perpendicular collisional transport in the drift-fluid equation gives rise to diffusion of particle density, momentum, and ion and electron pressure, and hence potentially renders the common use of artificial diffusion terms in turbulence models redundant. 

The collisonal terms are complex functions of the fluid and the electromagnetic field variables. Thus when included in turbulence models, they place significant demands on computing powers. Therefore, we also present a partly linearized model for the collisional transport across the magnetic field which allows significantly faster computing times.  The model is embedded in the hot edge-SOL electrostatic (HESEL) model\cite{rasmussen_EPS_2015,Nielsen2015}, the successor of the ESEL model\cite{Garcia_POP_062309,Funda_NF_2005_ESEL,militello_esel_2013,Yan_ESEL_2013}. HESEL is a two-dimensional (2D) model describing interchange turbulence in a plane perpendicular to the magnetic field at the outboard midplane of a tokamak. It evolves the particle density, the vorticity, and the electron and ion pressures. We show that our partly linearised collision terms, including inter species energy exchange, do not alter the energy theorem, and we show by means of 1D numerical simulations without turbulence that the simplified and full models give very similar results. 

The article is organized as follows. In section \ref{sec:PerpCollTrans} we describe how collisional effects influence the dynamics perpendicular to the magnetic field in a low-frequency drift-fluid turbulence model. We derive the global energy theorem and discuss energy exchange mechanisms due to collisions. In section \ref{sec:HeselModel} a simplified model for collisional effects is derived which is included in the HESEL model also presented here. Numerical simulations of the full and simplified models for the perpendicular collisional effects are presented, and finally in Sec.~\ref{sec:conclusions} we summarise our findings and draw conclusions.

\section{Perpendicular collisional transport in drift-fluid models}\label{sec:PerpCollTrans}
In this section we show how collisional transport across the magnetic field is consistently incorporated into drift-fluid models. Our derivation is based on the Braginskii\cite{Braginskii} fluid equations describing the time evolution of particle density $n_a$, velocity $ \bm{u}_a$ and temperature $T_a$ for a collisional, quasi-neutral $ n_i\simeq n_e$, electrostatic, simple electron-ion plasma. The index $a\in (i,e)$ is a species label, which refers to ions and electrons, respectively. We start our derivation from the Braginskii momentum equation 
\begin{align}
 n_am_a [\pfrac{}{t} + \bm{u}_a\cdot \nabla]\bm{u}_a = -\nabla p_a - \nabla \cdot \bm{\pi}_a + q_an_a(\bm{E} + \bm{u}_a\times \bm{B}) + \bm{R}_a.
 \label{eq:momentum}
\end{align} 
Here, $m_a$ denotes mass, $q_a$ is charge, $p_a= n_aT_a$ is the scalar pressure, $\bm{\pi}_a$ is the stress tensor, and $\bm{R}_a$ denotes the resistive force. The resistivity in the electron momentum equation consists of a frictional force 
\begin{align}
    \bm{R}_{e,u} = m_e n_e \nu_{ei} \big[0.51 (u_{\|i} - u_{\|e})\bhat + \bm{u}_{\perp i}-\bm{u}_{\perp e} \big]
        \label{eq:Ru}
\end{align}
and a thermal force 
\begin{align}
    \bm{R}_{e,T} = -0.71 n_e \bhat\nabla_{\|} T_e -\frac{3n_e \nu_{ei}}{2\Omega_e} \bhat \times \nabla T_e,
    \label{eq:RT}
\end{align}
where the electron-ion collision frequency is defined as
\begin{align}
    \nu_{ei}=\frac{\sqrt{2}}{12 \pi^{3/2} }\frac{  Z^2 e^4 \ln \Lambda}{\sqrt{m_e}  \epsilon_0^2}\frac{n_e}{T_e^{3/2}},
\end{align}
the magnitude of the electron gyrofrequency is $\Omega_e = \frac{e B}{m_e}$, and $\ln \Lambda$ denotes the Coulomb logarithm. We also introduced a unit vector parallel to the magnetic field $\bhat = \bm{B}/B$ and the magnetic field aligned component of the gradient operator $\nabla_{\|} = \bhat \cdot \nabla$. If $\bm{v}$ is an arbitrary vector, then we introduced the notation $v_{\|} = \bhat \cdot \bm{v}$ for the projection onto the magnetic field unit vector $\bhat$, and $\bm{v}_{\perp} = - \bhat \times (\bhat \times \bm{v})$ for the perpendicular part of the vector. Momentum conservation implies  $\bm{R}_{i} = - \bm{R}_{e}$. The origin of the thermal force $\bm{R}_{e,T}$ is the velocity dependence  of the particle collision frequency ($\propto v^{-3}$). The stress tensor $\bm{\pi}_a$ consists of three parts: i) A part $\bm{\pi}_a^{\|}$ describing viscosity along the magnetic field due to like-particle collisons,  ii) a gyro-frequency dependent part $\bm{\pi}_a^{\perp}$ describing collisional momentum transport across the magnetic field due to like particle collisions, and iii) a gyro-viscous part $\bm{\pi}_a^*$. Here we consider the $\bm{\pi}_i^{\perp}$ and $\bm{\pi}_i^*$ parts of the ion stress tensor. The electron stress tensor is neglected because it is smaller than the ion contributions by the electron-ion mass ratio. In the ion stress tensor we assume that the perpendicular flow is incompressible and assume a constant magnetic field. The pure perpendicular ion viscous tensor\cite{Braginskii} written in a local coordinate system $(x,y,z)$, where $\hat{\bm{z}}$ is aligned with $\bhat$, then reads: 
\begin{subequations}
\begin{align}
  \pi^{\perp}_{xx} &= -\pi^{\perp}_{yy}= -\eta^i_1 (\partial_x u_x- \partial_y u_y),\\
  \pi^{\perp}_{xy} &=\pi^{\perp}_{yx} = - \eta^i_1(\partial_x u_y + \partial_y u_x),   
\end{align}
  \label{eq:visc_perp_coll}
\end{subequations}
here the ion viscosity coefficient is given as
\begin{align}
    \eta_1^i = \frac{3}{10}\frac{ p_i \nu_{ii}}{ \Omega_i^2},
\end{align}
and the ion-ion collision frequency is 
\begin{align}
    \nu_{ii} = \frac{1}{12 \pi^{3/2}}\frac{ Z^4 e^4 \ln \Lambda }{m_i^{1/2} \epsilon_0^2}\frac{n_i}{T_i^{3/2}}.
\end{align}
The parallel-perpendicular parts, e.g., $\bm{\pi}^{\perp}_{xz}$, do not contribute to the perpendicular dynamics to lowest order and are therefore not treated here since we do not consider the dynamics parallel to the magnetic field. In the parallel direction the parallel-perpendicular part of the stress tensor give rise to perpendicular viscosity. 

In the drift ordering \cite{Hinton_Horton_1971} the fundamental assumptions are that the characteristic frequencies are much smaller than the ion gyrofrequency
\begin{align}
    \frac{\omega}{\Omega_i} \ll 1
\end{align}
and that perpendicular gradient length scales $L_{\perp}$ are larger than the ion gyro-radius 
\begin{align}
	\frac{\rho_i^2}{L_{\perp}^2}\ll 1.
	\label{eq:DF_orderingII}
\end{align}
We further assume $L_{\perp} \ll L_{\|}$, here $L_{\|}$ denotes a characteristic parallel gradient length scale.  This ordering permits iterative solutions  of the momentum equations for the perpendicular fluid drifts. The first order drifts are given as 
\begin{align}
    \bm{u}_{a\perp,1} &= \frac{\bhat \times \nabla \phi}{B} 
	      + \frac{\bhat \times \nabla p_a}{q_an_aB}	      
	  = \bm{u}_E + \bm{u}_{D a} 	 
	  \label{eq:vperp0}
\end{align}
which are the \ExB\ and diamagnetic drifts, respectively. The second order drifts are
\begin{align}
    \bm{u}_{a\perp,2} &= \Omega_a^{-1}\bhat \times \frac{d}{dt} \bm{u}_{a\perp,1}
	      - \frac{\bhat \times \bm{R}_a}{q_an_aB}
	      + \frac{\bhat \times \nabla \cdot \bm{\pi}_a}{q_an_aB}
		  = \bm{u}_{p a} + \bm{u}_{R	}+ \bm{u}_{\pi a}. 
		  \label{eq:vperp1}
\end{align}
In all second order drift we take $\bm{u}_{a\perp} = \bm{u}_{a\perp,1} $. The first term $\bm{u}_{a,p}$ is the polarization drift. Due to the mass dependence only the ion polarization drift is retained. The second term $\bm{u}_R$ is the drift\cite{HintonHazRevModPhys1976} associated with resistivity, which is identical for electrons and ions due to momentum conservation. The resistive drift consists of a friction force drift and a thermal force drift, respectively: 
\begin{align}
    \bm{u}_{R_u} &= -\frac{\nu_{ei}}{n_e m_e \Omega_e^2} \nabla_{\perp}P, \label{eq:uRu}\\
    \bm{u}_{R_T} &= \frac{3}{2}\frac{\nu_{ei}}{m_e \Omega_e^2}\nabla_{\perp}T_e.
    \label{eq:uRT}
\end{align}
Here $P=p_e + p_i$ denotes the total scalar pressure. Note that only the diamagnetic current appears in the friction force drift $\bm{u}_{R_u}$ because the electron and ion \ExB-drift contributions cancel. The diamagnetic drift is the fluid representation of the particle grad-B and curvature drifts, and the magnetization current associated with the gyrating charged particles\citep{Garcia_diamagnetic_fluid_part}. Therefore, the diamagnetic drift does not represent transport of guiding-centers unless the magnetic field is inhomogeneous. In that sense the friction force drift $\bm{u}_{R_u}$  is a fluid representation of friction between electrons and ions in opposite directed Larmor orbits. The direction of thermal force drift $\bm{u}_{R_T}$ is parallel to the electron temperature gradient. As already mentioned, this drift arises because fast particles experience less collisions than slow particles, which result in unbalanced fluxes, and hence implies up-gradient transport. The last first order drift is the viscous drift $\bm{u}_{\pi a}$ . As earlier mentioned this term is mass dependent and hence only the ion drift is retained. In the local coordinate system $(x,y,z)$ the purely perpendicular parts can be written as 
\begin{align}
 \bm{u}_{\pi i} = 
      \frac{1}{q_inB} \begin{pmatrix}
                       \partial_y \pi_{xx} - \partial_y \pi_{xy}\\
                       \partial_x \pi_{xx} + \partial_y \pi_{xy}\\
                       0
                      \end{pmatrix}.
                      \label{eq:u_pi}
\end{align}
Being a second order drift only the first order drifts $\bm{u}_{\perp,1}$, see Eq.~\reff{eq:vperp0}, are retained in the stress tensor $\bm{\pi}$.

\subsection{Collisional effects in the drift-fluid moment equations} \label{sec:moments}
With the algebraic expressions for the perpendicular drifts at hand, we can write down the resulting drift-fluid equations. We will omit the magnetic field aligned parts of the momentum equations since we are here mainly concerned with perpendicular dynamics. By inserting the perpendicular drifts given in equations \reff{eq:vperp0} and \reff{eq:vperp1} into the Braginskii\cite{Braginskii} particle density and pressure equations, we get: 
\begin{subequations}
\begin{align}
  \pfrac{}{t}n 
	    + \nabla \cdot (n\bm{u}_E) 
	    + \nabla \cdot (n\bm{u}_{De}) 
	    + \nabla \cdot (n\bm{u}_R)
	    + \nabla \cdot (\bhat u_{\|e}) =0	     \label{eq:df_n}\\
	       \nabla\cdot (n\bm{u}_{pi})
    +\nabla \cdot \big( n (\bm{u}_{Di}-\bm{u}_{De})\big)
    + \nabla \cdot (\bhat J_{\|}/e)
    + \nabla \cdot (n \bm{u}_{\pi_i}) = 0\label{eq:df_w}\\
      \frac{3}{2}\pfrac{}{t}p_e
      + \frac{3}{2}\nabla \cdot \big(p_e[\bm{u}_E + \bm{u}_{De} + \bhat u_{\|e} +\bm{u}_R ]\big)
      +p_e \nabla \cdot [\bm{u}_E + \bm{u}_{De} + \bhat u_{\|e}+\bm{u}_R ] \notag   \\
      + \nabla \cdot \bm{q}_{e} 
      + \nabla_{\perp} \cdot \bm{q}_{e}^*
= -Q_{\Delta} + \bm{R} \cdot (\bm{u}_i-\bm{u}_e)
\label{eq:df_pe}\\
    \frac{3}{2}\pfrac{}{t}p_i
      + \frac{3}{2}\nabla \cdot \big(p_i[\bm{u}_E + \bm{u}_{Di} + \bhat u_{\|i} + \bm{u}_{pi} + \bm{u}_{\pi i} + \bm{u}_R]\big)\notag \\
      +p_i \nabla \cdot [\bm{u}_E + \bm{u}_{Di} + \bhat u_{\|i} + \bm{u}_{pi} + \bm{u}_{\pi i} +\bm{u}_R] 
      + \nabla \cdot \bm{q}_{ i} 
      + \nabla_{\perp} \cdot \bm{q}_{\perp i}^*
      + \bm{\pi}_i^{\perp} : \nabla \bm{u}_{i\perp,1} 
= Q_{\Delta}. 
\label{eq:df_pi}
\end{align} 
\end{subequations}
Equation \reff{eq:df_w} is the vorticity equation which results from subtracting the ion and electron particle density equations assuming quasi-neutrality. In all equations higher order terms are retained even though some of these terms are formally small. The motivation for not neglecting these higher order terms is that they, e.g., guarantee energy conservation, provide diffusion or are responsible for  energy exchange between different plasma species. Drift-fluid equations have been derived by several authors\cite{Zeiler, Hinton_Horton_1971, DrakePOF1984, Simakov2003} but without self-consistently including perpendicular collisional transport. In the remainder of this section we describe and discuss the collisional terms entering the drift-fluid moment equations. Thorough descriptions and discussions of classical transport outside  the drift-fluid context is found in e.g. Refs.~\onlinecite{Braginskii,helander2005collisional,HintonHazRevModPhys1976}.

We start with the electron particle density equation \reff{eq:df_n} where the divergence of the resistive flux is the only collisional term. Using equations \reff{eq:uRu} and \reff{eq:uRT}, the divergence of the resistive flux can be written as
\begin{align}
    \nabla \cdot (n\bm{u}_R)	
     &= -\nabla \cdot \bigg(\nu_{ei} \rho_e^2 \bigg[
     (1 + \frac{T_i}{T_e})\np n
     + \frac{n}{T_e}(\np T_i-\frac{1}{2}\np T_e)\bigg] \bigg),
\label{eq:ResFlux}
\end{align}
demonstrating the origin of the resistive drifts, namely a perpendicular random-walk process with step length $\rho_e = \sqrt{T_e/(m_e \Omega_e^2)}$ and frequency $\nu_{ei}$. We see that the perpendicular friction force $\bm{R}_{u\perp}$ gives rise to particle density diffusion with a diffusion coefficient $\nu_{ei} \rho_e^ 2$ in drift-fluid models, similar to the explicit appearance of spatial diffusion in low-frequency ordered collision operators\cite{xu:1991,madsen2013gyrokinetic}. The thermal force drift $\bm{u}_{R_T}$ is partly canceled by the electron temperature gradient dependent part of the friction force drift $\bm{u}_{R_u}$, leaving only the last term on the right hand side. In order to highlight the $n$ and $T_e$ dependency of the electron-ion collision frequency $\nu_{ei} \propto n/T_e^ {3/2}$ we define 
\begin{align}
    \nu_{ei0}&=\nu_{ei} \frac{T_e^{3/2}}{n} \frac{n_0}{T_{e0}^{3/2}},
\end{align}
where $n_0$ and $T_{e0}$ denote a constant reference particle density and a constant reference electron temperature, respectively. Equation \reff{eq:ResFlux} then reads:
\begin{align}
    \nabla \cdot (n\bm{u}_R)	
      &= - \nabla  \cdot \bigg(\frac{\nu_{ei0}}{m_e \Omega_e^2}\frac{T_{e0}^{3/2}}{n_0} \frac{n}{\sqrt{T_e}}\bigg[
      	  (1 + \frac{T_i}{T_e}) \nabla_{\perp} n
	+ \frac{n}{T_e} (\nabla_{\perp}T_i -\frac{1}{2}\np T_e )
	 \bigg]\bigg),
\label{eq:ResFluxII}
\end{align}
showing that the particle density diffusion coefficient has a $(1+T_i/T_e)n/\sqrt{T_e}$ dependence. 

Next, consider the electron pressure equation \reff{eq:df_pe}. Here, collisional effects enter through classical heat fluxes, heat transfer terms, and resistive drift terms. The collisional electron heat flux\cite{Braginskii} consists of two parts:
\begin{align}
  \bm{q}_{e,u} &= -0.71 p_e (u_{i\|}-u_{e\|})\bhat   
	      - \frac{3}{2} \frac{p_e \nu_{ei}}{\Omega_e} \bhat \times (\bm{u}_{1, i \perp}-\bm{u}_{1, e \perp})\label{eq:electron_heatflux_qu},\\
  \bm{q}_{e,T} &= -\kappa_{e,\|} \bhat\nabla_{\|} T_e	
		  - \kappa_{e,\perp}\np T_e
		  \label{eq:electron_heatflux_qT}
\end{align}
where the heat conductivities\cite{Braginskii} are 
\begin{align}
  \kappa_{e,\|} = 3.16 \frac{p_e}{m_e \nu_{ei}}, \quad 
  \kappa_{e,\perp} = 4.66 n\frac{\nu_{ei}T_e}{m_e\Omega_e^2},
\end{align}
and where only the first order perpendicular drifts given in Eq.\reff{eq:vperp0} were used in $\bm{q}_{e,u}$. The origin of the heat flux $\bm{q}_{e,u}$ is the same as the thermal force $\bm{R}_{ei,T}$ given in Eq.~\reff{eq:RT}, which are both associated with the velocity dependence of the electron-ion collision frequency. Note the exact Onsager\cite{helander2005collisional} symmetric coefficients in $\bm{R}_{ei,T}$ and $\bm{q}_{e,u}$. It is convenient to express the perpendicular parts of the electron heat flux in terms of the perpendicular resistive drifts given in equations \reff{eq:uRu} and \reff{eq:uRT} 
\begin{align}
	\bm{q}_{e,u \perp} = - \frac{3}{2} p_e \bm{u}_{R_u}, \quad
	\bm{q}_{e,T \perp} = - \frac{28}{9} p_e \bm{u}_{R_T}. 
	\label{eq:q_u_rel}
\end{align} 
When written in this form it is evident that the divergence of the frictional electron pressure flux cancels the perpendicular part of $\bm{q}_{e,u}$:
\begin{align}
 \frac{3}{2} \nabla \cdot (p_e \bm{u}_{R_u}) + \np \cdot \bm{q}_{e,u \perp}= 0. 
 \label{eq:electron_heatflux_cancel}
\end{align}
On the right hand side of the electron pressure equation \reff{eq:df_pe}, the $ \bm{R} \cdot (\bm{u}_{1, i \perp}-\bm{u}_{1, e \perp})$ term transfers energy between the electron and ion fluid kinetic energies an the electron thermal energy. In the drift-fluid electron pressure equation the energy transfer term can be combined with the remaining perpendicular heat flux terms and terms which depend on the resistive drift
\begin{align}
    &\frac{3}{2}\nabla \cdot (p_e \bm{u}_{R_T})
    + p_e \nabla \cdot \bm{u}_R
    + \nabla \cdot \bm{q}_{e,T \perp}
    -  \bm{R} \cdot (\bm{u}_{i \perp,1}-\bm{u}_{e \perp,1})\notag  \\  
    &=
    \nabla \cdot (p_e \bm{u}_{R_u}) 
    -\frac{11}{18}\nabla \cdot (p_e \bm{u}_{R_T})
    +\bm{u}_R \cdot \np p_i. 
    \label{eq:pe_diff}
\end{align}
The two first terms on the right hand side, 
\begin{align}
    \nabla \cdot (p_e \bm{u}_{R_u}) 
    -\frac{11}{18}\nabla \cdot (p_e \bm{u}_{R_T})
    &= -\nabla \cdot \bigg(\nu_{ei} \rho_e^2 \bigg[\np P  + 
    \frac{11}{12} n\np T_e \bigg] \bigg),  
    \label{eq:pe_diffII}  
\end{align}
among others, yield diffusion of electron pressure and in general relax gradients, and hence the up-gradient flux contained in $\bm{u}_{R_{T,\perp}}$ is canceled by perpendicular heat conduction in $\bm{q}_{e,T}$.  The last term in Eq.~\reff{eq:pe_diff} transfers energy between the electron and ion thermal energy due to resistivity. Hence, no explicit energy transfer channel between kinetic and thermal energy due to resistivity remains in the drift-fluid equations. Energy exchange between comoving electron and ion fluids is described by the heat exchange term 
\begin{align}
  Q_{\Delta} = 3\frac{m_e}{m_i} n\nu_{ei}(T_e - T_i)
  \label{eq:energy_equilibration}
\end{align}
entering the electron and ion pressure equations \reff{eq:df_pe} and \reff{eq:df_pi}, respectively. The equilibration occurs on the slower $\nu^{-1}_{ei} m_i/m_e$ collision time-scale because the energy transfer in each scattering event is proportional to the mass ratio of the scattering particles $m_e / m_i$. We elaborate further on inter-species energy exchange in section \ref{sec:energyconservation}.

We now turn to the ion pressure equation \reff{eq:df_pi} where the heat conduction is given as 
\begin{align}
    \bm{q}_{i} &= -\kappa_{i,\|} \nabla_{\|} T_i	
		  - \kappa_{i,\perp}\np T_i
\label{eq:ion_heatflux}		  
\end{align}
and the thermal conductivities\cite{Braginskii} are
\begin{align}
    \kappa_{i,\|} = 3.9 \frac{p_i}{m_i \nu_{ii}}, \quad 
  \kappa_{i,\perp} = 2 n\nu_{ii}\rho_i^2.
  \label{eq:electronheatcond}
\end{align}
The ion heat flux is solely driven by ion temperature gradients and is independent of ion-electron collisions. The ratio between the perpendicular electron and ion heat conductivities is of the order $ \kappa_{e,\perp}/\kappa_{i,\perp} \sim \sqrt{m_e/m_i}$ for $T_e \sim T_i$, whereas $ \kappa_{e,\|}/\kappa_{i,\|} \sim \sqrt{m_i/m_e}$. Perpendicular collisional heat transport is therefore  dominated by the ions, and parallel heat transport is electron dominated. The terms involving the resistive drift in the ion pressure equation can be written as
\begin{align}    
    \frac{3}{2}\nabla \cdot (p_i \bm{u}_R)    
    + p_i \nabla \cdot \bm{u}_R    
    = \frac{5}{2}\nabla \cdot (p_i \bm{u}_R) - \bm{u}_R \cdot \np p_i\notag \\ 
    =  \frac{5}{2}\nabla \cdot \bigg(\frac{T_i}{T_e}\nu_{ei} \rho_e^2 \big[\frac{3}{2}n\np T_e - \np P\big]\bigg) - \bm{u}_R \cdot \np p_i,
    \label{eq:pi_diff}
\end{align}
which shows that electron-ion collisions through the resistive drift give rise to ion pressure diffusion, but also that the thermal force drift $\bm{u}_{R_T}$ potentially drives the ion pressure up the electron temperature gradient. Note that the corresponding effect in the electron pressure equation was canceled by the electron heat conduction. The diffusion coefficient is proportional to the electron diffusion coefficient times the ratio of the ion to the electron temperature. The ratio of the resistive diffusion and the perpendicular ion heat conduction can be estimated as 
\begin{align}
	\frac{|\nabla \cdot \bm{q}_{i\perp}|}{|\nabla \cdot (p_i \bm{u}_R)|}
	\sim \sqrt{\frac{m_i}{m_e}} \sqrt{ \frac{T_e}{T_i}} \frac{L_{T_e}^ 2}{L_{T_i}^2},
	\label{eq:ion_cond_vs_resi}
\end{align}
where $L_{T_e}$ and $L_{T_i}$ are characteristic electron and ion temperature gradient length scales. Perpendicular heat conduction therefore dominates resistive diffusion except for plasmas where the ion temperature is significantly higher than the electron temperature or where the electron temperature has much steeper gradients than the ion temperature. 

In both pressure equations the diamagnetic heat flux is defined as 
\begin{align}
	\bm{q}_a^* = \frac{5}{2}\frac{p_a}{q_aB} \bhat \times \nabla T_a.
\end{align}

Lastly, we consider viscous effects in the vorticity and ion pressure equations \reff{eq:df_w} and \reff{eq:df_pi}, respectively. In both equations viscous effects enter through the viscous drift $\bm{u}_{\pi}$ defined in Eq.~\reff{eq:vperp1}, but the viscous tensor only explicitly appears in the ion pressure equation. Notice that only the divergence of the viscous particle density flux remains in the vorticity equation. The resistive electron and ion fluxes cancel as a consequence of momentum conservation. 
The divergence of the perpendicular viscous flux results in diffusion of the magnetic field aligned \ExB\ and ion diamagnetic vorticities due to ion-ion collisions, as we will show in section \ref{sec:HeselModel}. In the ion pressure equation the corresponding ion pressure flux gives rise to hyperviscosity "$\propto -\nabla^4 p_i$" of the ion pressure. However, compared with the perpendicular ion heat conduction the viscous ion pressure flux term is in most cases small 
\begin{align}
  \frac{| \nabla \cdot (p_i \bm{u}_{\pi i})| }{|\nabla \cdot \bm{q}_{i,\perp}|}
\sim \frac{\rho_i^2}{L^2_{\perp}}. 
  \label{eq:viscpiflux_to_qi}
\end{align}
The viscous drift and the viscous tensor provide an energy transfer channel between ion thermal energy and kinetic energy through the $p_i \nabla \cdot \bm{u}_{\pi i}$ and the $\bm{\pi}_i^{\perp} : \nabla \bm{u}_{i\perp}$ terms. The underlying energy transfer mechanism is the randomization of ordered perpendicular fluid motion due to ion-ion collision. As we will show in the next section, both terms are important for the conservation of energy. The gyro-viscous part vanishes\cite{Smolyakov1998} exactly $\bm{\pi}_i^{*} : \nabla \bm{u}_{i\perp,1} = 0$, when the gyro-viscous tensor $\bm{\pi}_i^{*}$ is evaluated with the first order drift velocities $\bm{u}_{i\perp,1}$.

\subsubsection{The polarization drift and the gyro-viscous cancellation} \label{sec:pola_gyr_canc}
In this section explicit expressions for the polarization and gyro-viscous drifts are given. These drifts enter the \textit{gyro-viscous} cancellation which eliminates the advection of vorticity by the diamagnetic drift in the polarization equation \reff{eq:df_w}, but also the advection of ion parallel momentum and ion heat fluxes by the diamagnetic drift in their corresponding moment  equations\cite{Smolyakov1998}. These equations are significantly altered by the gyro-viscous tensor. The gyro-viscous cancellation is not complete in the sense that numerous small correction terms remains in the vorticity equation, in the parallel momentum equation, and in the heat flux equation. The actual derivation is cumbersome and the level of complication rises with the level of detail included in the polarization drift terms, e.g., the inclusion of the polarization heat flux and anisotropic pressure\cite{Smolyakov1998,Belova_2001}. 

Here, we aim at formulating a workable model for the use in numerical turbulence simulations. Therefore, we do not include the polarization heat flux and we do not account for anisotropic pressures. Furthermore, we bring the magnetic unit vector under the material derivative in the polarization drift $\Omega^{-1}\bhat \times \frac{d}{dt} \bm{u}_{i1,\perp} \simeq\Omega^{-1}\frac{d}{dt} \bhat \times \bm{u}_{i1,\perp}$ and neglect the corresponding correction terms in the gyro-viscous drift. This approximation yields simpler expressions compared to the full expressions in e.g. Refs.~\onlinecite{Zeiler,Belova_2001,Smolyakov1998}. As already discussed, we leave out parallel momentum for the purpose of exposition. After carrying out the gyro-viscous cancellation\cite{Smolyakov1998,Belova_2001}, the combined divergences of the ion polarization and gyro-viscous fluxes are: 
\begin{align} 
	\nabla \cdot (n\bm{u}_{p i} + n\bm{u}_{\pi^* i}) 
	\simeq -\nabla \cdot \bigg(\frac{n}{\Omega_i} d_t^ \# \np\Phi^* \bigg) 	
	+\nabla \cdot (n\bm{u}_{\chi i}),
	\label{eq:GV_cancel}
\end{align}
where we introduced the short-hand notation 
\begin{align}
	\np \Phi^* \doteq \frac{\np \phi}{B}  + \frac{\np p_i}{qnB}
\end{align}
and where the material derivative is defined as  $d_t^\# =\partial_t  + \bm{u}^\# \cdot \nabla$. The advecting velocities are
\begin{align}
	\bm{u}^\# =  
	 \bm{u}_{i, 1\perp} + \bm{u}_{p i} + \bm{u}_{\chi i} + \bm{u}_{R} + \bm{u}_{\pi^{\perp}_i}.
\end{align}
Here the drift 
\begin{align}
	\bm{u}_{\chi i} = -\frac{T_i}{B} \curl{\frac{\bhat }{q_iB}} \cdot \nabla \np\Phi^*
\end{align}
represents the remainder of the gyro-viscous cancellation. $\bm{u}_{\chi i}$ is derived from the scalar function $\tilde{\chi}$ given in Eq.~37 in Ref.~\cite{Belova_2001}. The scalar function $\tilde{\chi}$ enters the momentum equation as a correction to the scalar pressure and hence only contributes when the magnetic field is inhomogeneous. The contributions originating from  $\tilde{\chi}$ are therefore small. Here, we choose to keep only the terms in $\tilde{\chi}$ necessary for energy conservation. Furthermore, terms in $\tilde{\chi}$ originating from the purely perpendicular heat flux are neglected because energy conservation requires that these contributions are retained together with the polarization heat flux which is not kept here. 

As a rule of thumb\cite{pop2003BDS} all ion drifts retained in the ion continuity equation, except the ion diamagnetic drift, must be kept in the advection part of the polarization drift in order to conserve energy. Unfortunately, this leads to a recursive definition of the polarization drift since the polarization drift includes a polarization drift advection term. If exact energy conservation is a crucial demand, this feature makes the model unsuitable for numerical calculations, but the model is convenient when discussing energy conservation in drift-fluid models in general.

\subsection{Energy conservation and energy exchange} \label{sec:energyconservation}
In this section we present the energy theorem for the drift-fluid model Eqs.~\reff{eq:df_n}-\reff{eq:df_pi}. Without taking collisional effects into account it has previously been shown, e.g., Refs.~\onlinecite{Zeiler,pop2003BDS}, that drift-fluid models conserve energy. Here, the drift-fluid energy theorem is presented including the perpendicular collisional transport terms  and the  collisional energy exchange terms described in section \ref{sec:moments}. 

The global energy is obtained by multiplying the vorticity equation \reff{eq:df_w} by $q_i\phi$ and integrate over all space neglecting surface terms 
\begin{align}
\int dV \, & -q_in(\bm{u}_{p i} + \bm{u}_{\chi i})\cdot \nabla \phi   
	-\np \phi \cdot [q_in (\bm{u}_{Di}-\bm{u}_{De})
    + q_i n  \bm{u}_{\pi^{\perp} i}]\notag \\
=    \int dV \, &nm_i \frac{\np \phi}{B} \cdot  d_t^\#  \np \Phi^*
+	p_i \curl{\frac{\bhat }{q_iB}} \cdot \nabla \np \Phi^* \cdot \frac{\np \phi}{B}
	-\bm{u}_E \cdot \nabla P
 + (\nabla \cdot \bm{\pi}^{\perp}_i) \cdot \bm{u}_E.
    \label{eq:Energy_kin}
\end{align}
Terms involving parallel dynamics are everywhere left out. Detailed descriptions of energy conservation and energy transfer channels in the parallel direction in the absence of perpendicular collisional effects is thoroughly described in Refs.~\onlinecite{Zeiler} and \onlinecite{pop2003BDS}. Next, the ion and electron pressure equations are integrated over space using Eqs.~\reff{eq:electron_heatflux_cancel}, \reff{eq:pe_diff}, and \reff{eq:pi_diff}, again neglecting surface terms
\begin{align}
	\int dV \, &  \frac{3}{2}\pfrac{}{t}p_e
      -\nabla p_e  \cdot \bm{u}_E 
       +\bm{u}_R \cdot \np p_i
+ Q_{\Delta} = 0, 
\label{eq:Energy_pe}\\
    \int dV \,& \frac{3}{2}\pfrac{}{t}p_i
+    nm_i \frac{\np p_i}{q_inB} \cdot  d_t^\#  \np \Phi^*
+	p_i \curl{\frac{\bhat }{q_iB}} \cdot \nabla \np \Phi^* \cdot \frac{\np p_i}{q_inB}\notag\\
     &- \nabla p_i \cdot \bm{u}_E  
	  + (\nabla \cdot \bm{\pi}_i^{\perp})\cdot \bm{u}_{Di}  
      -\bm{u}_R \cdot \np p_i
      + \bm{\pi}_i^{\perp} : \nabla \bm{u}_{i\perp,1} 
 - Q_{\Delta} = 0,
 \label{eq:Energy_pi}
\end{align}
where $\bm{u}_{pi}$ and $\bm{u}_{\chi i}$ were inserted into the $p_i \nabla \cdot (\bm{u}_{pi} + \bm{u}_{\chi i})$ term, resulting in the second and third terms on the right hand side of Eq.~\reff{eq:Energy_pi}, respectively. The energy theorem 
\begin{align}
\frac{d}{dt}	\int dV \, \frac{3}{2}p_e + \frac{3}{2}p_i + \frac{1}{2}m_i n u_{i\perp,1}^2 = 0 
\label{eq:cons_energy}
\end{align}
is obtained by adding Eqs.~\reff{eq:Energy_kin}-\reff{eq:Energy_pi} and by bringing  the ion particle density under the time derivative in terms involving the ion polarization drift using the particle density equation \reff{eq:df_n} and the vorticity equation \reff{eq:df_w}. As expected, the energy theorem is not altered by the inclusion of perpendicular collisional effects. In other words, the collisional effects do not give rise to sinks no sources in the energy theorem, but solely provide energy transfer channels. In that respect the energy theorem is a sanity check of the model and can be used as a verification tool in numerical simulations. 

The perpendicular kinetic energy $\frac{1}{2}m_i n u_{i\perp,1}^2$ is a peculiar quantity. The \ExB-drift $\frac{1}{2}m_i n u_E^2$ part is easy to interpret, because the fluid moments in  Eqs.~\reff{eq:df_n}-\reff{eq:df_pi} are all advected by the \ExB-drift, and hence this part describes the kinetic energy associated with the \ExB-drift. On the contrary, the terms in the kinetic energy $\frac{1}{2}m_i n u_{i\perp,1}^2$ involving the diamagnetic drift are more difficult to interpret because the diamagnetic drift to lowest order does not advect fluid elements. The diamagnetic part of the kinetic energy is more appropriately described as a finite Larmor radius (FLR) correction to the ion pressure and to the ion \ExB-kinetic energy\cite{scott:102318,jmad2011FLRBlob,wiesenberger2014}. To emphasize this property of the kinetic energy, it is denoted the \textit{modified kinetic energy}.

To elucidate how energy is transferred between the constituents of the energy theorem Eq.~\reff{eq:cons_energy} it is instructive to consider the time evolution of each term separately
\begin{alignat}{2}
&\frac{d}{dt} \int dV\, \frac{1}{2}nm_i u_{i\perp,1}^2 
&&= \int dV \, -P \nabla \cdot \bm{u}_E
 + p_i \nabla \cdot (\bm{u}_{pi} + \bm{u}_{\chi i})
 -(\nabla\cdot \bm{\pi}_i^{\perp}) \cdot \bm{u}_E,
 \label{eq:ddt_Ekin}
 \\
 &\frac{d}{dt} \int dV\, \frac{3}{2}p_i 
  &&=\int dV\, p_i \nabla \cdot \bm{u}_E
 -p_i \nabla \cdot (\bm{u}_{p i} + \bm{u}_{\chi i})
 + (\nabla\cdot \bm{\pi}_i^{\perp}) \cdot \bm{u}_E
 +\bm{u}_R\cdot \nabla p_i 
  +Q_{\Delta},
   \label{eq:ddt_Epi}
  \\
 &\frac{d}{dt}  \int dV\,  \frac{3}{2} p_e
 &&=\int dV\, p_e \nabla \cdot \bm{u}_E
 -\bm{u}_R\cdot \nabla p_i 
 -Q_{\Delta}.
    \label{eq:ddt_Epe} 
\end{alignat}
Terms which enter with opposite signs represent conservative energy transfer channels.  The interchange transfer terms, first terms on the right hand sides of Eqs.~\reff{eq:ddt_Ekin}-\reff{eq:ddt_Epe}, transfer energy between ion and electron thermal energies and the modified kinetic energy. The compression of the polarization drift, second term on the right hand sides of Eqs.~\reff{eq:ddt_Ekin} and \reff{eq:ddt_Epi}, provides an energy transfer between ion thermal energy and the modified kinetic energy through the $p_i \nabla \cdot (\bm{u}_{p} + \bm{u}_{\chi})$ terms. This energy transfer channel may play an important role in generating and sustaining \ExB \ mean flows\cite{ReynoldsJMAD2015}. 

The heat exchange term $Q_{\Delta}$, entering Eqs.~\reff{eq:ddt_Epe} and \reff{eq:ddt_Epi} with opposite signs, describes energy exchange between electrons and ions due to elastic electron-ion collisions. Resistivity provides an additional energy transfer channel through the resistive drift transfer terms $\bm{u}_R \cdot \np p_i$ entering Eqs.~\reff{eq:ddt_Epe} and \reff{eq:ddt_Epi} with opposite signs.  The parallel resistivity, on the contrary, gives rise to an energy exchange between the ion and electron parallel kinetic energies $\frac{1}{2}m_an u_{a\|}^2$ and the electron thermal energy. Here, resistivity opposes the motion of the charge carrying particles along the magnetic field. In the perpendicular direction, resistivity opposes the diamagnetic current, which to lowest order describes particle gyration around magnetic field lines. Since the kinetic energy of the gyrating motion is included in the thermal energies, perpendicular resistivity in the drift-fluid model accordingly give rise to a conservative energy between the ion and electron thermal energies. Note that a finite resistive energy transfer requires a finite ion pressure gradient. When $\np p_i = 0 $, ions are homogeneously distributed and therefore the probability of an electron colliding with a gyrating ion is the same in all directions. Only when $\np p_i \neq 0$, electrons do on average experience an average effect of the gyrating ions. The direction of the energy transfer can be deduced by writing out the transfer term in the electron pressure equation explicitly
\begin{align}
\frac{3}{2}\pfrac{}{t}p_e + \cdots =     -\bm{u}_R \cdot \np p_i
    &=-\nu_{ei}\rho_e^ 2 \bigg[\frac{1}{2} \frac{\np T_e}{T_e} - \bigg(1+\frac{T_i}{T_e}\bigg)  \frac{\np n}{n} - \frac{\np T_i}{T_e} \bigg] \cdot \np p_i,      
    \label{eq:R_E_trans}
\end{align}
where the thermal force and the electron temperature gradient dependent part of the friction force were combined. The second and third terms on the right hand side heat the electrons whenever $\np n \cdot \np T_i > 0 $. The heating of electrons due to friction is known as Joule heating. However, it can be shown that the second and third terms are also capable of cooling the electrons. Necessary conditions are $\np n \cdot \np T_i < 0 $ and that the ion temperature gradient length scale exceeds the particle density gradient length scale. The electron temperature gradient dependent term transfers thermal energy from the electron to the ions when $\np T_e \cdot \np p_i > 0$, an effect known as the Ettingshausen effect\cite{Hinton1984}. 

It is important to keep in mind that both the heat exchange term $Q_{\Delta}$ and the resistive transfer terms occur on the slow time-scale $\nu_{ei}^{-1} m_i/m_e$. Comparing the magnitudes of these energy exchange terms 
\begin{align}
    \frac{|Q_{\Delta} |}{| \bm{u}_R \cdot \np p_i |}
    \sim \frac{T_e -T_i}{T_e} \frac{L_{\perp}^2}{\rho_i^2}
\end{align}
shows that unless the temperatures are very different, the resistive energy transfer is smaller than the heat exchange when considering profiles, but also that the resistive can be important when pressure gradients are steep, e.g., in the edge region and in general when turbulent transport is strong. 

Viscosity on the perpendicular flows due to ion-ion collisions gives rise to the energy transfer term $(\nabla\cdot \bm{\pi}^{\perp}) \cdot \bm{u}_E,$ between the modified kinetic energy and the ion thermal energy. We should expect that the energy transfer goes from kinetic energy to thermal energy because collisions increase entropy. If the perpendicular electric field vanishes the energy transfer term vanishes accordingly. It can be shown that the $\bm{u}_E$ dependent part of $\bm{\pi}^{\perp}$ entering the transfer term gives rise to an unidirectional energy transfer from modified kinetic energy to the ion thermal energy. The direction of the energy transfer by the remaining part, which depends on the ion diamagnetic drift, is not unidirectional. However, recall that the modified kinetic energy does not solely describe kinetic energy. Therefore, in order to investigate the direction of the energy transfer one must instead consider the energy transfer from the (FLR) corrected ion \ExB \ kinetic energy. Disentangling the true kinetic energy from the modified kinetic energy is beyond the scope of this work.

\section{The HESEL model}\label{sec:HeselModel}
Turbulent transport in the SOL region is intermittent and is carried by coherent structures\cite{blobreviewMyra}. These coherent structures of elevated pressure, often referred to as filaments or blobs, are elongated along the magnetic field. They are mainly born in the vicinity of the last closed flux surface (LCFS) at the outboard midplane due to ballooning\cite{Gunn2007}. Once these plasma outbreaks are expelled into the the open field line region, they expand\cite{MoultonPPCV2013} into the rare SOL plasma in the parallel direction with velocities comparable to the ion sound speed $c_s = \sqrt{(T_e+T_i)/m_i}$, while traversing the SOL region radially at velocities reaching a fraction of $c_s$.  Typically, the blobs imply fluctuation levels which exceed unity. 

In this section we derive a reduced model for investigations of blob formation and transport across the SOL region of magnetically confined toroidal plasmas. The model is named HESEL\cite{rasmussen_EPS_2015,Nielsen2015} and is the successor of the ESEL model\cite{Garcia_POP_062309,Funda_NF_2005_ESEL,militello_esel_2013,Yan_ESEL_2013}. 
Like the ESEL model, the HESEL model considers the dynamics in a plane perpendicular to the magnetic field located at the outboard midplane. Perpendicular collisional transport is described by partly linearized expressions derived from the more general results presented in the preceding section. Parallel transport in the SOL region is represented by damping terms. Parallel drift-wave dynamics and magnetic perturbations are ignored. Turbulent fluctuations are hence driven by the interchange instability due to an inhomogenious toroidal magnetic field. These reductions facilitate long but also computationally inexpensive computer simulations which can provide sufficient data for statistical analysis and comparison with experiment. Compared to the ESEL model the HESEL model has the following enhancements: 1) A dynamical description of ion pressure dynamics, 2) the inertial response to changes in the ion pressure is included, 3) sheath dissipation, and 4) an improved description of collisional processes, including inter-species energy exchange, based on the results presented in the previous section. 

The model domain is a plane perpendicular to the magnetic field at the outboard midplane. The magnetic field geometry is approximated by local slab coordinates $(x,y,z)$. $x$ and $y$ coordinates denote the radial and the poloidal azimuthal positions in the domain and $z$ denotes the position along the magnetic field. The toroidal magnetic field magnitude is approximated as $B = B_0(1 + \epsilon + x /R) $, where $\epsilon = r/R$ denotes the inverse aspect ratio, and $r$ and $R$ are the minor and major radii, respectively. The HESEL model consists of two regions joined at the LCFS ($x = 0$) modeling the edge and SOL regions. Besides radial boundary conditions, these regions are only set apart by damping terms due to losses along open field lines which are only active in the SOL region. In this section we present and discuss the simplifications of the drift-fluid model equations \reff{eq:df_n}-\reff{eq:df_pi} that lead to the HESEL model.
 
\subsection{Thin-layer approximation} \label{sec:thinlayer}
First we turn our attention to the vorticity equation \reff{eq:df_w}. Here we invoke an approximation commonly known as the thin-layer approximation. This approximation shares many features with the Boussinesq approximation often used in the description of incompressible, buoyancy-driven flows in neutral fluid dynamics, but is fundamentally different since the mechanism driving the dynamics in plasma drift-fluid equations is finite compressibility of the plasma fluid drifts. This rather crude approximation neglects particle density variations in the polarization flux entering the vorticity equation \reff{eq:df_w}, and hence assumes a constant inertia of all fluid parcels irrespective of the local particle density. The motivation for this approximation, which in particular is debatable when applied to edge plasma models, is the desire for computational efficiency and that energy conservation requires polarization drift advection in the polarization drift itself as discused in section \ref{sec:pola_gyr_canc}. This recursive definition is very inconvenient in numerical computations. We note that gyrofluid models\cite{madsen2013GF} do not suffer from these difficulties. 

Specifically, we approximate the ion polarization flux defined in Eq.~\reff{eq:GV_cancel} as 
\begin{align} 
 \nabla \cdot (n[\bm{u}_{p i}+ \bm{u}_{\chi i}])=
		 -\nabla \cdot \bigg(\frac{n}{\Omega_i} d_t^ \# \np\Phi^* \bigg) 	
	+\nabla \cdot (n\bm{u}_{\chi})
	\simeq 	 -\frac{n_0}{\Omega_{i0}} \nabla \cdot (d_t^0 \np\phi^*\big), 	
	\label{eq:HESEL_TL}
\end{align}
where we have introduced 
\begin{align}
	\phi^* = \frac{\phi}{B_0} + \frac{p_i}{q_in_0B_0}
	\label{eq:HESEL_modpot}
\end{align}	
and where subscripts "$0$" on  quantities on the right hand side refer to  constant characteristic reference values. The material derivative
\begin{align}
	\frac{d^0}{dt} = \pfrac{}{t} + \bm{u}_{E0} \cdot \nabla
\end{align} 
only includes the \ExB-drift at constant magnetic field $\bm{u}_{E0} = B_0^{-1} \bhat \times \nabla \phi$. Dictated by energy conservation, all other advection terms in the original polarization flux are neglected.  

Energy conservation requires that the divergence of the polarization drift in the ion pressure equation \reff{eq:df_pi} is approximated in the same way
\begin{align} 
	p_i \nabla \cdot (\bm{u}_{p i}+ \bm{u}_{\chi i})
	\simeq -\frac{p_i}{\Omega_{i0}} \nabla \cdot (d_t^0 \np\phi^*\big).  	
	\label{eq:pol_flux_thinlayer}
\end{align}
In the final equations the divergence of the polarization drift does not enter the ion pressure equation, because the vorticity equation is substituted into the ion pressure equation. The divergence of the polarization flux $\frac{3}{2} \nabla \cdot (p_i [\bm{u}_{p i}+ \bm{u}_{\chi i}])$ is formally small and does not affect energy conservation. It is therefore neglected.

\subsection{Approximations for the perpendicular collisional dynamics}
We start by considering the particle density equation \reff{eq:df_n} where the only perpendicular collisional term is the divergence of the resistive flux written out in Eq.~\reff{eq:ResFluxII}. As previously noted, the divergence of the resistive flux gives rise to particle density diffusion with a diffusion coefficient $D_e\propto n /\sqrt{T_e}$. Naturally,  the particle density and electron temperature fields in the edge and SOL regions of a tokamak plasma will not be completely correlated. However, we expect that the profiles of both fields share features such as being flat in the SOL region and having steep gradients near the separatrix. Furthermore, measurements in the SOL region at the outboard midplane show that $n$ and $T_e$ are approximately proportional in blobs\cite{Nold_2012} in which case $D_e \propto \sqrt{n}$. Therefore, we choose to fix the diffusion coefficient in Eq.~\reff{eq:ResFluxII} evaluating it using characteristic constant reference values $n_0$ and $T_{e0}$ for the particle density and the electron temperature, respectively. Recall that the thermal force was partly canceled leaving only the last term in Eq.~\reff{eq:ResFluxII}. This residual thermal force is expected to be partly canceled by the ion temperature gradient part whenever the ion and electron temperatures are correlated. Therefore, we neglect the temperature gradient terms in the resistive flux Eq.~\reff{eq:ResFluxII}. The divergence of the perpendicular resistive flux in the particle density equation \reff{eq:df_n} then reads
\begin{align}
    \nabla \cdot (n \bm{u}_{R0})
    \simeq -D_e (1+\tau) \nabla_{\perp}^2 n,
    \label{eq:ur0}
\end{align} 
where $D_e = \nu_{ei0} \rho_{e0}^2 $ is the constant diffusion coefficient calculated using $n_0$ and $T_{e0}$, and $\tau = T_{i0}/T_{e0}$ is the reference ion to electron temperature ratio. Usage of the same approximate resistive drift  
\begin{align}
	\bm{u}_{R0} = -D_e (1+\tau) \np \ln n
\end{align}
in both the electron and ion pressure equations \reff{eq:df_pe}-\reff{eq:df_pi} avoids undesired temperature dynamics such as temperature anti-diffusion. 

Next we consider the electron pressure equation \reff{eq:df_pe}. As was shown in Eq.~\reff{eq:pe_diff}, the resistive fluxes and the heat fluxes partially cancel. Apart from the energy exchange terms, the remaining terms were given in Eq.~\reff{eq:pe_diffII}. These terms are approximated as 
\begin{align}
 \nabla \cdot (p_e \bm{u}_{R_{u,\perp}}) 
    -\frac{11}{18}\nabla \cdot (p_e \bm{u}_{R_{T,\perp}})
    &= -\nabla \cdot \bigg(\nu_{ei} \rho_e^2 \bigg[\np P  + 
    \frac{11}{12} n\np T_e \bigg] \bigg)    \notag  \\
    &\simeq \nabla \cdot (p_e \bm{u}_{R0}) 
- \nabla \cdot \bigg(D_e \frac{11}{12} n\np T_e \bigg] \bigg)  .   
\end{align}
The resistive energy transfer terms entering the electron and ion pressure equations are approximated as
\begin{align}
  \bm{u}_{R} \cdot \np p_i 
  \simeq
  \bm{u}_{R0} \cdot \np p_i.
\end{align}

In the ion pressure equation \reff{eq:df_pi} the collisional transport has a different character. Here, the perpendicular ion heat conduction dominate the divergence of the resistive flux. However, the resitive flux is retained in order to handle situations with flat ion temperature gradients properly, see Eq.~\reff{eq:ion_cond_vs_resi}. As in the electron pressure equation the divergence of the ion pressure resistive flux is approximated as 
\begin{align}
	\frac{5}{2}\nabla \cdot (p_i \bm{u}_R) 
	\simeq 	\frac{5}{2}\nabla \cdot (p_i \bm{u}_{R0}).
\end{align}
The perpendicular ion heat conduction Eq.~\reff{eq:ion_heatflux} 
\begin{align}
      \bm{q}_{i\perp} \simeq 2 n D_i \np T_i
\end{align}
is approximated by taking the ion diffusion coefficient 
\begin{align}
D_i = \frac{\nu_{ii,0} T_{i0}}{m_i \Omega_{i0}^2}
\label{eq:iondiffusioncoef}
\end{align} 
to be constant.

In the electron and ion pressure equations \reff{eq:df_pe}-\reff{eq:df_pi} the heat exchange term defined in Eq.~\reff{eq:energy_equilibration} is evaluated with a constant electron-ion collision frequency 
\begin{align}
	Q_{\Delta} \simeq  3\frac{m_e}{m_i} n\nu_{ei0}(T_e - T_i).
\end{align}

Finally, we consider approximations of terms related to the perpendicular collisional ion viscosity tensor $\bm{\pi}_i^{\perp}$, given in equation \reff{eq:visc_perp_coll}, which enters the vorticity and ion pressure equations. In the vorticity equation \reff{eq:df_w}, $\bm{\pi}_i^{\perp}$ gives rise to diffusion of vorticity via the divergence of the flux associated with the viscous drift $\bm{u}_{\pi^{\perp}_i}$. Recall that in order to reduce the computational costs, the thin-layer approximation to the polarization flux was invoked as described in section \ref{sec:thinlayer}. The thin-layer approximation implied that the first order perpendicular ion drifts entering the polarization drift itself were approximated as
\begin{align}
	\bm{u}_{i \perp,1 } \simeq 
		\bm{u}^0_{i \perp,1} = \bhat \times \np \phi^*.
		\label{eq:uperpzero}
\end{align}
Accordingly, only $\bm{u}^0_{i \perp,1}$ enters $\bm{\pi}_i^{\perp}$. Expressed in  local coordinates $(x,y,z)$ we get 
\begin{align}
	  \pi^{\perp}_{xx,0} &= -\pi^{\perp}_{yy,0}= 2\eta_1^i\partial^2_{xy} \phi^*,\\
  \pi^{\perp}_{xy,0} &=\pi^{\perp}_{yx,0} = - \eta_1(\partial^2_{xx} -  \partial^2_{yy} )\phi^*. 
\end{align}
Furthermore, the thin-layer approximation takes the particle density in the ion polarization flux to be constant. Making the same approximation to the viscous drift flux, we get 
\begin{align}
     \nabla \cdot (n \bm{u}_{\pi^{\perp}_i} )
     \simeq    \nabla \cdot (n_0 \bm{u}^0_{\pi^{\perp}_i} )
	  = \frac{3}{10}\frac{n_0 D_i}{\Omega_{i0}}\nabla_{\perp}^2 \nabla_{\perp}^2 \phi^*,
	  \label{eq:viscosity}
\end{align}
which demonstrates that the perpendicular viscous drift gives rise to diffusion of the modified vorticity $\omega^* = \np^2 \phi^*$. Vorticity diffuses due to ion-ion collisions which randomize the ions and hence transfer energy from the modified kinetic energy, see Eq.~\reff{eq:ddt_Ekin}, to the ion thermal energy through the terms  $p_i \nabla \cdot \bm{u}_{\pi^{\perp}_i}$ and $\bm{\pi}_i^{\perp}: \nabla \bm{u}_{i \perp,1 }$ entering the ion pressure equation \reff{eq:df_pi}. Energy conservation requires that the perpendicular drifts entering $\bm{\pi}_i^{\perp}$ and $\bm{u}_{\pi^{\perp}_i}$ are approximated according to equation \reff{eq:uperpzero}: 
\begin{align}
  \bm{\pi}_i^{\perp} : \nabla \bm{u}_{i\perp,1}
  \simeq   \bm{\pi}_{i}^{\perp 0} : \nabla \bm{u}^0_{i \perp,1}
  &=
      -\frac{3}{10} m_i n_0D_i\bigg[ 
      (\partial_{xx}\phi^* - \partial_{yy}\phi^*)^2 + 4 (\partial_{xy} \phi^*)^2
      \bigg],\\
      p_i\nabla \cdot \bm{u}_{\pi^{\perp}} 
      \simeq p_i\nabla \cdot \bm{u}^0_{\pi^{\perp}_i} 
	  &= p_i\frac{3}{10}\frac{D_i}{\Omega_{i0}}\nabla_{\perp}^2 \nabla_{\perp}^2 \phi^*.
	  \label{eq:visc_flux_comp_pi}
\end{align}
The $\nabla \cdot (p_i \bm{u}_{\pi^{\perp}})$ term in the ion pressure equation is formally small, see Eq.~\reff{eq:viscpiflux_to_qi}. This term does not change the energy theorem and is therefore neglected.

\subsubsection{Neoclassical corrections}\label{sec:neo}
In a toroidal plasma perpendicular collisional diffusion is enhanced by neoclassical transport. Specifically, in a collisional plasma, $\lambda/L_{\|} \ll 1$, the transport along the magnetic field is diffusive; here $\lambda$ denotes the mean free path and $L_{\|}$ is the parallel length scale. Despite being interrupted by collisions, guiding-centers drifts, such as the grad-B drift, give rise to transport across magnetic flux surfaces. In a tokamak, the direction, inwards or outwards, depends on whether particles are above or below the outboard midplane. When the diffusive transport along the magnetic field and the perpendicular guiding center drifts are superimposed they give rise to an enhanced Pfirsch-Schl\"uter\cite{PfirschS1962,HintonHazRevModPhys1976,helander2005collisional}  perpendicular diffusion $D_{nc}  \propto \textsl{q}^2$, where $\textsl{q}$ is the safety factor.   

Neoclassical transport theory assumes that the plasma evolves on a time-scale much slower than the typical time-scale of turbulent transport. Therefore, neoclassical estimates of perpendicular transport do not immediately apply to intermittent transport in the SOL region.  Furthermore, existing literature is only concerned with neoclassical transport on closed field lines. Here, we are interested in local transport on closed \textit{and} open field-lines. However, it is evident that the physical mechanism driving Pfirsch-Schl\"uter transport is also present on open field lines and that it will influence profiles as well as transients, but since a rigorous derivation of these effects is beyond the scope of this work, we resort to the existing estimates keeping in mind that these will only hold approximately. Specifically, diffusion coefficients are multiplied by a common factor  
\begin{align}
	D_{e,i} \rightarrow (1+  \frac{R}{a}\textsl{q}_{95}^2) D_{e,i} 
\end{align}	
which then include variations of the inverse aspect ratio $\epsilon = a/R$ at the outboard midplane. Since the safety factor becomes infinite at the LCFS and is not defined on open field lines we estimate the effect of a finite magnetic field line pitch by the safety factor at $95\% $ of the poloidal magnetic flux $\textsl{q}_{\text{95}}$. A more detailed discussion is given in Fundamenskii et al Ref.~\onlinecite{Funda_NF_2005_ESEL}. It is important to note that since the perpendicular diffusive particle flux is enhanced by neoclassical effects, so is the resistive energy transfer, second last term in Eq.~\reff{eq:ddt_Epe}, between the electron and ion thermal energies\cite{HintonHazRevModPhys1976}.

\subsection{Parallel losses in the SOL region}
The HESEL model is 2D and therefore parallel losses on open field lines are parametrized. The parametrization is in line with the parametrization in ESEL model given in  Ref.~\onlinecite{Funda_NF_2005_ESEL}, but it is augmented by parametrization of parallel ion pressure dynamics and the addition of a damping term representing sheath currents at material surfaces such as divertors and limiters. As already mentioned, the primary aim of the HESEL model is to describe interchange driven turbulence transport at the outboard midplane. As a consequence of ballooning the turbulent plasma source mainly resides on the high field side at the outboard midplane. For this reason, we assume that all parametrized parallel terms in the HESEL model act as sinks. 

Blobs expand\cite{MoultonPPCV2013} in the parallel direction with a velocity comparable to the ion sound speed $c_s = \sqrt{( T_e+ T_i)/m_i}$. The expansion depletes the particle density in the model-domain located at the outboard midplane. Under the assumption that both ends of the blob expand with the same velocity, we get the characteristic parallel particle density damping rate
\begin{align}
  \frac{1}{\tau_{n}} = \frac{2 M c_s}{L_{b}},
\end{align}
where $M = u_{\|i}/c_s$ denotes the Mach number and $L_b$ is the parallel blob size. We assume that blobs are predominantly born on the high field side in a region centred around the outboard midplane with a  $60^{\unit{\circ}}$ poloidal extend\cite{Gunn2007}, and hence approximate the filament blob size as 
\begin{align} 
	L_b = \frac{2\pi \textsl{q}_{\text{95}} R}{6} \simeq \textsl{q}_{\text{95}} R.
\end{align}

In the SOL region vorticity losses are due to two mechanisms: First, parallel advection of vorticity enters through the divergence of the ion polarization drift particle density flux $\nabla \cdot (n\bm{u}_{pi}) = - u_{i\|}\zhat \cdot \nabla \np^2\phi^* + \cdots $. Since vorticity is mainly transported and generated by blobs in the SOL region, we assume that the parallel gradient length scales of vorticity and particle density are approximately the same. In that case, the damping rate parametrising parallel advection of vorticity is then taken to be identical to the particle density damping rate: 
\begin{align}
    \frac{1}{\tau_w} =     \frac{1}{\tau_n}. 
\end{align} 
Secondly, the parallel current in the vorticity equation provides an additional path for current density generated by the diamagnetic current in blobs. The role played by the parallel current is roughly determined by the parallel conductivity, the magnetic topology particularly in the vicinity of the x-point, the sheath in front of material surfaces e.g. the divertor in most tokamaks, and the ratio of the characteristic time $t_{\perp}$ it takes a blob to move a distance in the radial direction comparable to its own size to the characteristic time $t_{\|}$ of the parallel communication of current perturbations between the outboard midplane and the sheath. Here, we consider blobs in the inertial regime where radial blob convection is barely inhibited by parallel currents. This regime is typically relevant for plasmas with a high particle density, a long connection length $L_{b}$ between the outboard midplane and the divertor, or a low conductivity e.g., due to high neutral gas densities in the SOL region\cite{blobreviewMyra}. Nevertheless, we assume that the divertor sheath influences the background profile of the electrostatic potential. An average in the parallel direction\cite{Krasheninnikov2001368} of the parallel current leaves us with the value of the current at the sheath entrance given by the Bohm criterion\cite{stangeby2000plasma}. For this reason we only apply the sheath damping to the averaged fields
\begin{align}
    \nabla \cdot (\bhat J_{\|}/e) \simeq 
    \mathcal{S} =
    \frac{en_0\GA{c_s}}{L_c}\bigg[1-\exp\big(\Lambda-\frac{e\GA{\phi}}{\GA{T_e}} \big)\bigg],
    \label{eq:sheathDamping}
\end{align}
where $\Lambda = \log(\sqrt{\frac{m_i}{2\pi m_e}})$ is the Bohm potential and the azimuthal average is defined as  
\begin{align} 
	\GA{f} = \frac{1}{L_y} \int_0^{L_y} dy\, f.
\end{align}
Here, $L_y$ is the domain size in the $y$-direction and $f$ is an arbitrary function. It is straightforward to use other closures of the parallel current e.g., the sheath-connected regime\cite{blobreviewMyra} can be investigated by applying the  sheath damping not only to the averaged fields but to the fluctuating parts as well. 

We now turn to the electron pressure equation \reff{eq:df_pe}. Here, the parallel heat fluxes given in Eqs.~\reff{eq:electron_heatflux_qu}-\reff{eq:electron_heatflux_qT} and the parallel advection term are parametrized. Quasi-neutrality only allows small deviations between the ion and electron parallel velocities. Consequently, the thermal heat flux $q_{e,u\|}$ given in Eq.~\reff{eq:electron_heatflux_qu} is therefore small in comparison with the divergence of the parallel pressure flux and is therefore neglected in the model. The Spitzer-H\"arm\cite{spitzerH_PRL1953} parallel electron heat conduction $q_{e,T\|}$ given Eq.~\reff{eq:electron_heatflux_qT}, on the other hand, is not negligible. It has a strong electron temperature dependence $\kappa_{e,\|} \propto T_e^{5/2}$, which attenuates parallel electron temperature gradients. Keep in mind that the Braginskii closure is only valid in the collisional regime. In particular $q_{e,T\|}$ is not well behaved as we approach the "collisionless" regime. This is readily seen when writing the heat conductivity in terms of the mean free path $\lambda_e$  
\begin{align}
    \nabla_{\|} \cdot \bm{q}_{e,T\|}
    \simeq -3.16 \, n T_e^{3/2}\frac{\lambda_e}{L_{\|}^2}.
\end{align}
As a rule of thumb the Braginskii heat conduction is therefore only valid as long as the electron collisionality is sufficiently high ($\nu_e^* > 10$). At lower collisionalities the Spitzer-H\"arm heat conduction exceeds the free-streaming heat flux\cite{fundamenskii_2005_heatflux}. A remedy for this shortcoming is to introduce \textit{heat-flux-limiters}, which in an ad-hoc manner limit the parallel heat flux to a fraction of the free-streaming value. Here we restrict ourselves to collisional SOL conditions where the Spitzer-H\"arm heat conductivity is valid. Conduction is then parametrized as 
\begin{align}
     \nabla_{\|} \cdot \bm{q}_{e,T\|}
     \simeq -\frac{T_e^{7/2}}{\tau_{\text{SH}}},
\end{align}
where we introduced the Spitzer-H\"arm damping rate 
\begin{align}
\tau_{\text{SH}}^{-1} =
	3.16 \frac{n_0}{m_e \nu_{ei0} T_{e0}^{3/2} L_c^2}. 
\end{align}
In the parametrization of the electron heat conduction we take the connection length $L_c$ as the parallel gradient scale length and not the parallel filament size $L_b$ to avoid an unphysical damping of the electron temperature background. Parallel advection of electron and ion pressure are approximated as a one-dimensional adiabatic expansion ($T \propto n^2$) 
\begin{align}
    \frac{1}{\tau_{p_e}} = 
    \frac{1}{\tau_{p_i}} 
    =\frac{9}{2} \frac{1}{\tau_n}.
\end{align}

In the ion pressure equation parallel heat conduction is neglected. For the parameters used here (ion collisionality $\nu_i^* > 10$ and $M > 0.1$, see  Fig.14 in Ref.\onlinecite{Funda_NF_2005}), parallel ion heat transport is dominated by parallel advection.

\subsection{Resulting model equations}
We are now ready to write down the full set of equations which constitute the HESEL model. In order to highlight characteristic quantities and for convenience, the model equations are gyro-Bohm normalized 
\begin{align}
	\Omega_{i0} t \rightarrow t, \quad
	\frac{x}{\rho_s} \rightarrow x, \quad
	\frac{T_{e,i}}{T_{e0}} \rightarrow T_{e,i},\quad
	\frac{e\phi}{T_{e0}} \rightarrow \phi, \quad
	\frac{n}{n_0}\rightarrow n,	
\end{align}
where $\Omega_{i0} = eB_0/m_i $ is the characteristic ion gyro frequency, $\rho_s = \sqrt{\frac{T_{e0}}{m_i \Omega_{i0}^2}}$ is the cold-ion hybrid thermal gyro-radius, $n_0$, $T_{e0}$, and $T_{i0}$ are characteristic reference values for the particle density and the temperatures. The full system of equations is:
\begin{subequations}
 \begin{align}
  	  \frac{d}{dt}n 	    
	    + n\mathcal{K}(\phi)
	    - \mathcal{K}(p_e)
	    &=\Lambda_n,\label{eq:n_norm}\\    
 \nabla \cdot \big(\frac{d^0}{dt}  \np \phi^*\big) 
    -  \mathcal{K}(p_e + p_i)    
     &= \Lambda_w,\label{eq:w_norm}
     \\
     \frac{3}{2}\frac{d}{dt}p_e
      + \frac{5}{2}p_e\mathcal{K}(\phi)
      -  \frac{5}{2}\mathcal{K}\big(\frac{p_e^2}{n}\big)                         
      &= \Lambda_{p_e}, 
\label{eq:electronpressure}\\
\frac{3}{2}\frac{d}{dt}p_i
      + \frac{5}{2}p_i\mathcal{K}(\phi)
      +  \frac{5}{2}\mathcal{K}\big(\frac{p_i^2}{n}\big)
      - p_i  \mathcal{K}(p_e + p_i)    
           &= \Lambda_{p_i},
\label{eq:ionpressure_thinlayer}
\end{align}
\end{subequations}
where the advective derivatives are defined as 
\begin{align}
	 \frac{d}{dt} = \pfrac{}{t} + B^{-1}\{\phi,\cdot\}, \quad 
	\frac{d^0}{dt} = \pfrac{}{t} + \{\phi,\cdot\}.
\end{align}
\ExB-advection is written in terms of the anti-symmetric bracket 
\begin{align}
	\{f,g\}  = \pfrac{f}{x}\pfrac{g}{y} - \pfrac{f}{y}\pfrac{g}{x}.
\end{align}
The curvature operator is defined as 
\begin{align}
  \mathcal{K}(f)  
  = -\frac{\rho_s}{R}\pfrac{}{y}f,
  \end{align}
and the normalized modified potential reads
\begin{align}
    \phi^* = \phi + p_i.
\end{align}
In the ion pressure equation, the divergences of the polarization and viscous drifts were eliminated by substitution of the vorticity equation. Diamagnetic advection in the electron and ion pressure equations is no longer present as a consequence of the diamagnetic cancellation 
\begin{align}
      \frac{3}{2}\nabla \cdot(p_a \bm{u}_{Da})
      +p_a\nabla \cdot(\bm{u}_{Da})
      + \nabla \cdot \bm{q}_{a}^* 
=  \frac{5}{2} \curl\bigg(\frac{\bhat }{q_aB}\bigg) \cdot \nabla (p_a T_a).
\end{align}
All terms related to diffusion and parallel damping are grouped on the right hand sides of Eqs.~\reff{eq:n_norm}-\reff{eq:ionpressure_thinlayer} and are given as 
\begin{align}
      \Lambda_n =&  D_e (1+\tau) \np^2 n
	     - \sigma(x)\frac{n}{\tau_n} \\
      \Lambda_w =&  \frac{3}{10} D_i \np^2 \np^2 \phi^*
     -\sigma(x)\frac{w}{\tau_n}     
     +\sigma(x)\mathcal{S}
		\\
      \Lambda_{p_e} =& 
      D_e (1+\tau)\nabla \cdot(  T_e \np n)
    	 + D_e \frac{11}{12} \nabla \cdot(  n\np T_e )
	    + D_e (1+\tau)\np \ln n \cdot \np p_i \notag \\
	 & -\frac{3 m_e}{m_i}\nu_{ei0}(p_e- p_i)  
	  - \sigma(x)\bigg[\frac{9}{2}\frac{p_e}{\tau_n} 
	 + \frac{T_e^{7/2}}{\tau_{\text{SH}}}\bigg]\\     
     \Lambda_{p_i} = 
	  &\frac{5}{2}D_e (1+\tau) \nabla \cdot (T_i\np n)
	-D_e (1+\tau)\np \ln n \cdot \np p_i 
	+ 2D_i\nabla \cdot (n\np T_i) \notag \\
	&\frac{3}{10}D_i \big[	 
      (\partial^2_{xx}\phi^* - \partial^2_{yy}\phi^*)^2 + 4 (\partial^2_{xy} \phi^*)^2
	\big]
	+\frac{3 m_e}{m_i}\nu_{ei0}(p_e- p_i)	
	+ \sigma(x)\bigg(
	 p_i \mathcal{S} 	
	- \frac{9}{2}\frac{p_i}{\tau_n}\bigg)	,
\end{align}
where $\tau = T_{i0}/T_{e0}$ and
\begin{align}
	\sigma(x) = 
		\frac{\sigma_s}{2}\bigg[ 1+ \tanh(\frac{x-x_s}{\delta_s})\bigg]
\end{align}
is a ``smooth step function'' which defines the transition from closed to open field line regions in the model. The normalized diffusion coefficients are
\begin{align}
	D_e = (1+\frac{R}{a} \textsl{q}_{\text{95}}^2)\frac{\rho_{e0}^2 \nu_{ei0}}{\rho_s^2 \Omega_{i0}}, \quad 
		D_i = (1+\frac{R}{a} \textsl{q}_{\text{95}}^2)\frac{\rho_{i0}^2 \nu_{ii0}}{\rho_s^2 \Omega_{i0}}
\end{align}
and the damping rates due to parallel advection and parallel Spitzer-H\" arm  heat conduction are 
\begin{align}
  \tau_{n}^{-1} = \frac{2 M \sqrt{\frac{ T_e+ T_i }{m_i}}}{\textsl{q}_{\text{95}}R} , \qquad
  \tau_{\text{SH}}^{-1} =
	3.16 \frac{n_0}{m_e \nu_{ei0} T_{e0}^{3/2} L_c^2},
\end{align}
where $M$ is the Mach number. The sheath damping term entering the vorticity and ion pressure equations is given as 
\begin{align}
    \mathcal{S} =
    \frac{en_0\GA{c_s}}{L_c}\bigg[1-\exp\big(\Lambda-\frac{e\GA{\phi}}{\GA{T_e}} \big)\bigg].
\end{align}

\subsubsection{Energy conservation}
The thin-layer approximation alters the conserved energy compared to conserved energy of the full drift-fluid model described in Sec.~\ref{sec:energyconservation}. The derivation paths are very similar. First, the pressure equations \reff{eq:electronpressure} and \reff{eq:ionpressure_thinlayer} are integrated neglecting surface terms. Next, the vorticity equation \reff{eq:w_norm} is multiplied by "$-\phi$" \reff{eq:w_norm} and integrated, again neglecting surface terms. Adding the results we obtain the energy  theorem 
\begin{align}
      \frac{d}{dt} \int d\bm{x} \,    \frac{|\np \phi^*|^2}{2} 
      + \frac{3}{2} [p_i + p_e]. 
      =\int d\bm{x}\, S_{\|}.
\end{align}
The energy sinks due to parallel losses are contained in 
\begin{align}
    S_{\|} = -\sigma(x)\bigg(\frac{\phi \np^2 \phi^*}{\tau_w} 
    + \frac{p_i}{\tau_{p_i}} 
    + \frac{p_e}{\tau_{p_e}}
	+ \phi^*\mathcal{S}\bigg).
\end{align}
As discussed in Sec.~\ref{sec:energyconservation} the electron and ion energy equilibration terms $Q_{\Delta}$, the                                                                                                                                                                                                                                                                                                                                                                                                                                                                                                                                                                                                                                                             frictional heat exchange terms $\bm{u}_R \cdot \np p_i$, and the viscous terms exactly cancel and therefore do not enter the energy theorem.  Just as in the energy theorem for the full model given in Eq.~\reff{eq:cons_energy}, collisions do not give rise to energy sinks and sources but merely give rise to conservative energy transfer terms in the model. A thorough discussion of energy transfer mechanisms and the impact on mean flows can be found in Refs. \onlinecite{MadsenPPCF2015,ReynoldsJMAD2015}.

\subsection{Testing the model for perpendicular collisional diffusion}
In order to test the simplified model for perpendicular collisional tranport applied in the HESEL model, we have solved a subset of the governing equations \reff{eq:n_norm}-\reff{eq:ionpressure_thinlayer} numerically using the BOUT++ framework\cite{bout2009}. Specifically, we assume that: 1) The system is homogeneous in the y-direction, 2) only terms describing collisional transport are retained, and 3) parallel sinks are not active i.e. $\sigma(x) = 0$. In this particular limit, the HESEL equations reduce to the 1D coupled set of non-linear equations
\begin{align}
	\left.
	\begin{aligned}
		\pfrac{}{t}n = \Lambda_n, \quad
		\frac{3}{2}\pfrac{}{t}p_e = \Lambda_{p_e}, \quad
		\frac{3}{2}\pfrac{}{t}p_i = \Lambda_{p_i}
	\end{aligned}
	\right\}\quad \text{HESEL.}
	\label{eq:HESEL}
\end{align}   

In order to validate the simplified model, we have solved the full collisional model in the same limit. The full model is described in details in section \ref{sec:moments}. In the absence of turbulence the model equations reduce to:  
\begin{align}
\left.
\begin{aligned}
	\pfrac{}{t}n +\nabla \cdot (n\bm{u}_R) &= 0 ,  \\
	\frac{3}{2}\pfrac{}{t}p_e 
	-\nabla \cdot \bigg(\nu_{ei} \rho_e^2 \bigg[\np P  + 
    \frac{11}{12} n\np T_e \bigg] \bigg) 
    +\bm{u}_R \cdot \np p_i&= -Q_{\Delta}, \\
	\frac{3}{2}\pfrac{}{t}p_i 
	+\nabla \cdot \bm{q}_{i\perp}
	+\frac{5}{2}\nabla \cdot (p_i \bm{u}_R)
		- \bm{u}_R \cdot \np p_i 		
	 &= Q_{\Delta}
	 \end{aligned}
	 \right\} \quad \text{Full Model (FM)}
	 \label{eq:FM}
\end{align}
The resistive drift $\bm{u}_R$ is given in Eqs.~\reff{eq:uRu}-\reff{eq:uRT}, the heat exchange term $Q_{\Delta}$ is given in Eq.~\reff{eq:energy_equilibration}, and the perpendicular ion heat flux $\bm{q}_{i\perp}$ is given in Eq.~\reff{eq:ion_heatflux}. We stress that no terms are linearised and that collision frequencies are evaluated as dynamical functions of temperatures and particle density. 

\begin{figure}[htbp]
\begin{minipage}[c]{.49\textwidth}
  \par\vspace{0pt}
  \centering 
\includegraphics[width=1\textwidth]{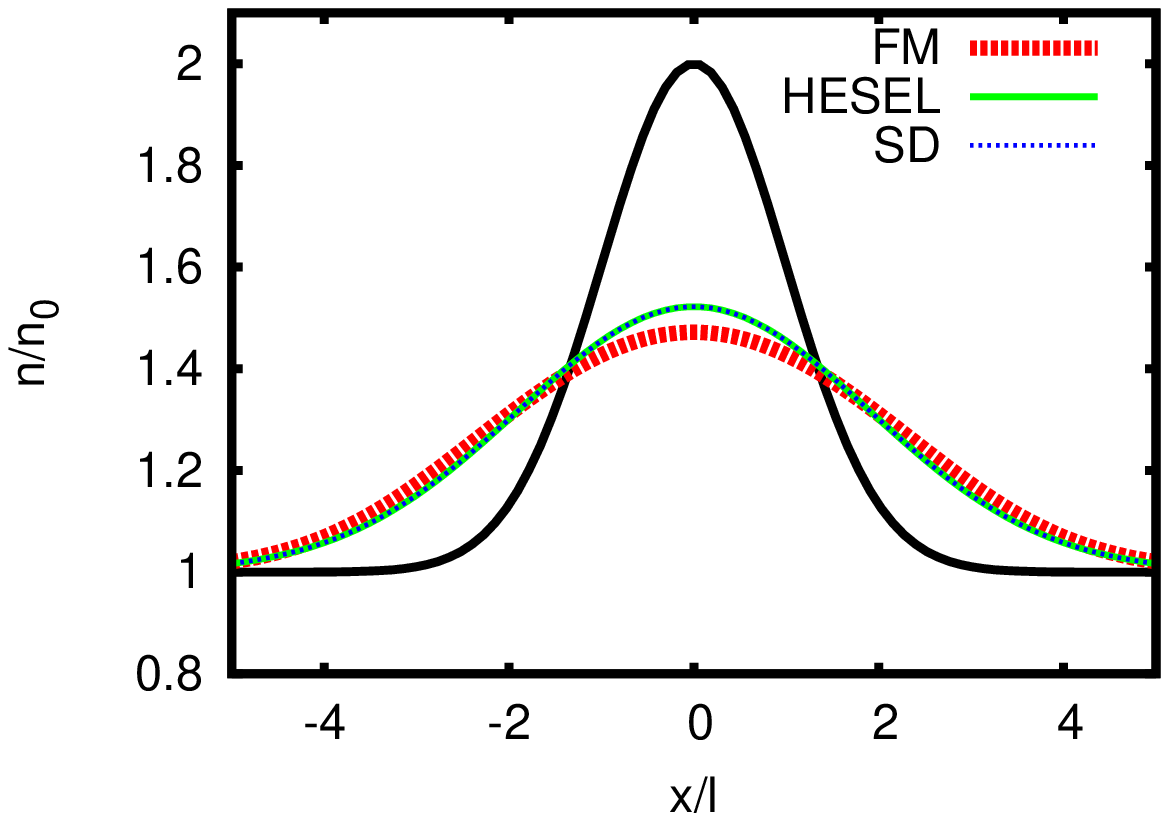}
\label{fig:n} 
\end{minipage}
\begin{minipage}[c]{.49\textwidth}
  \par\vspace{0pt}
  \centering 
\includegraphics[width=1\textwidth]{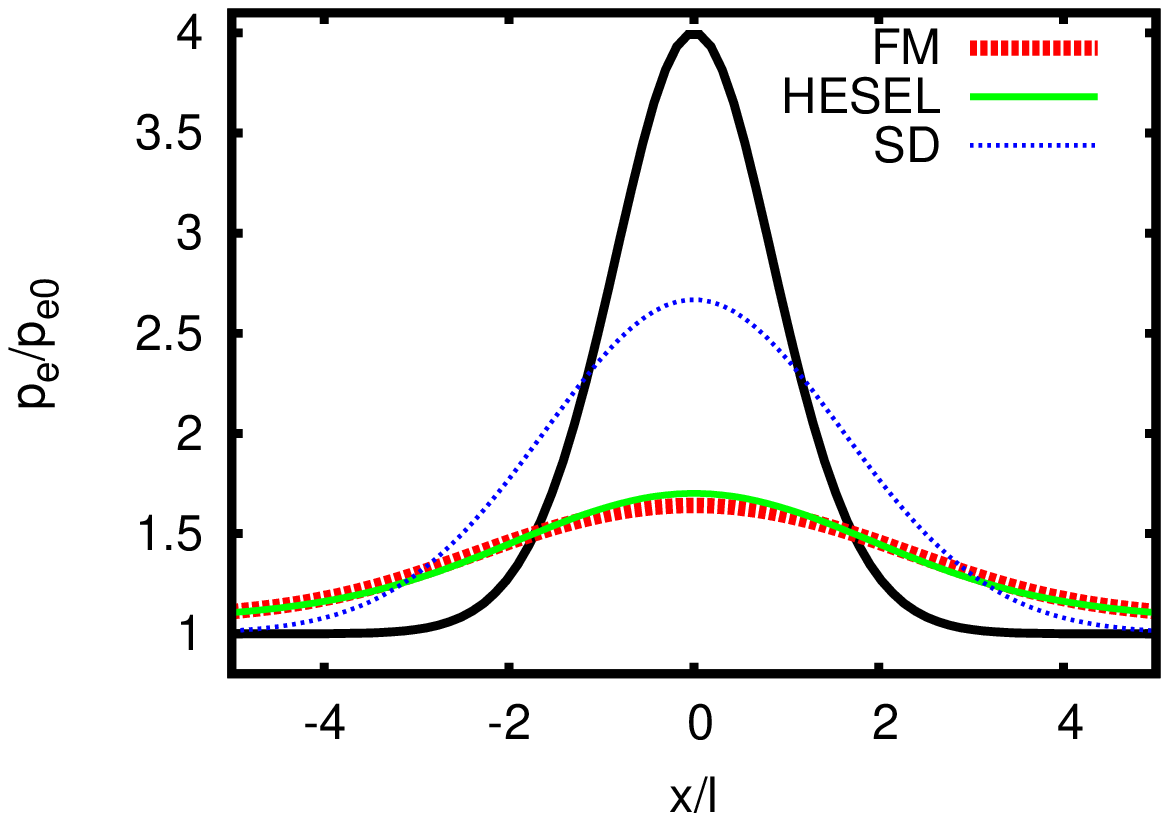}
\label{fig:pe} 
\end{minipage}
\begin{minipage}[c]{.49\textwidth}
  \par\vspace{0pt}
  \centering  
\includegraphics[width=1\textwidth]{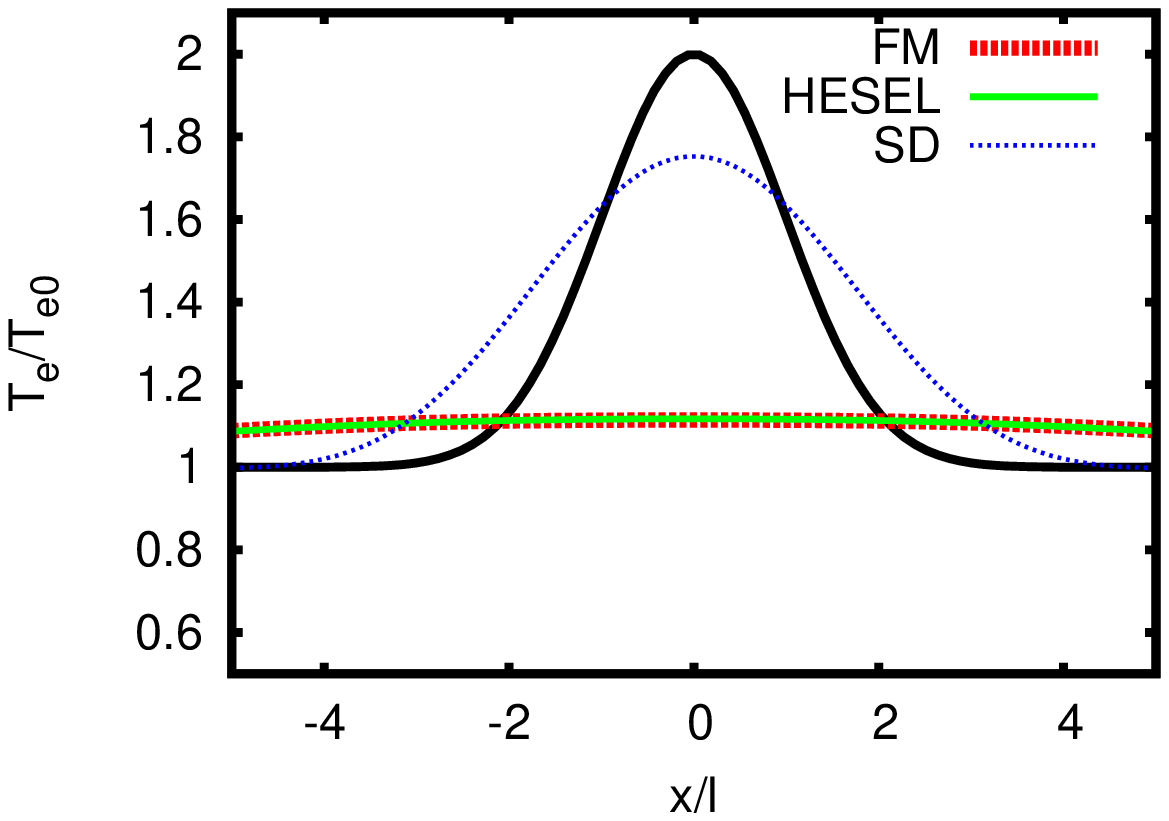}
\label{fig:te} 
\end{minipage}
\begin{minipage}[c]{.49\textwidth}
  \par\vspace{0pt}
  \centering 
\includegraphics[width=1\textwidth]{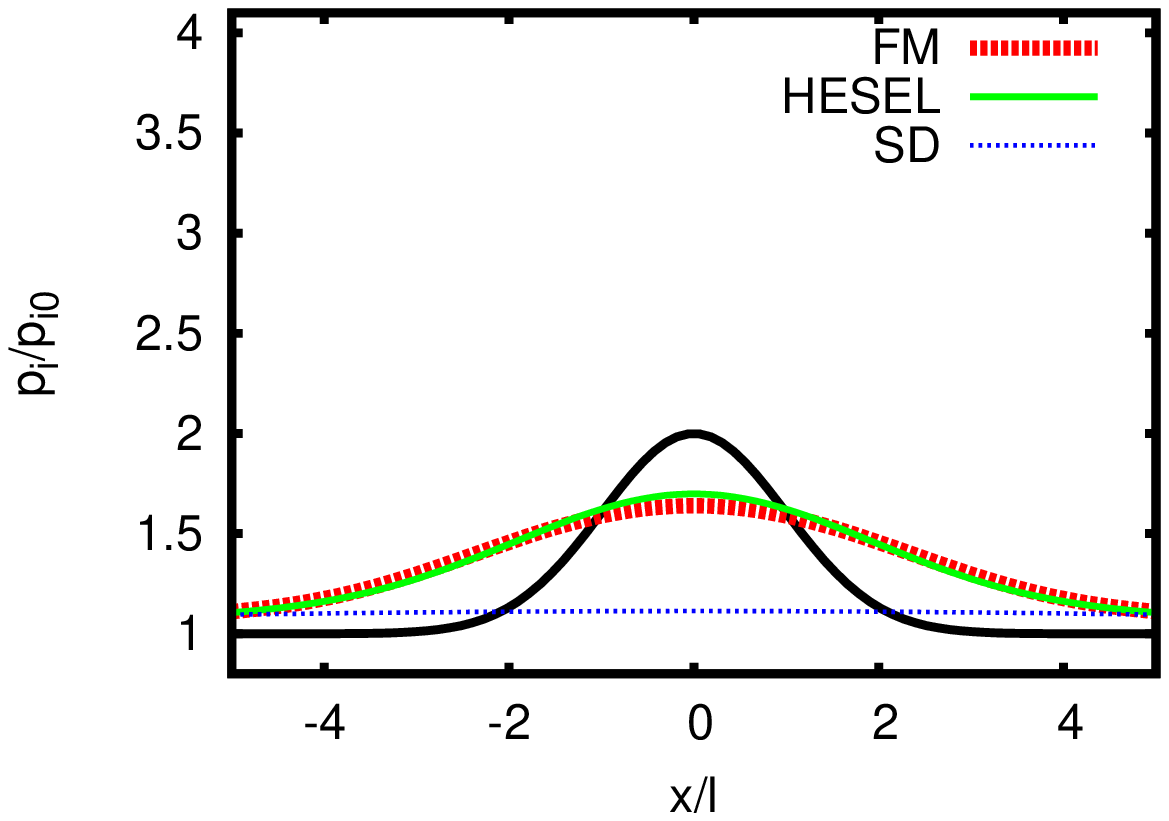}
\label{fig:pi} 
\end{minipage}
\begin{minipage}[c]{.49\textwidth}
  \par\vspace{0pt}
  \centering 
\includegraphics[width=1\textwidth]{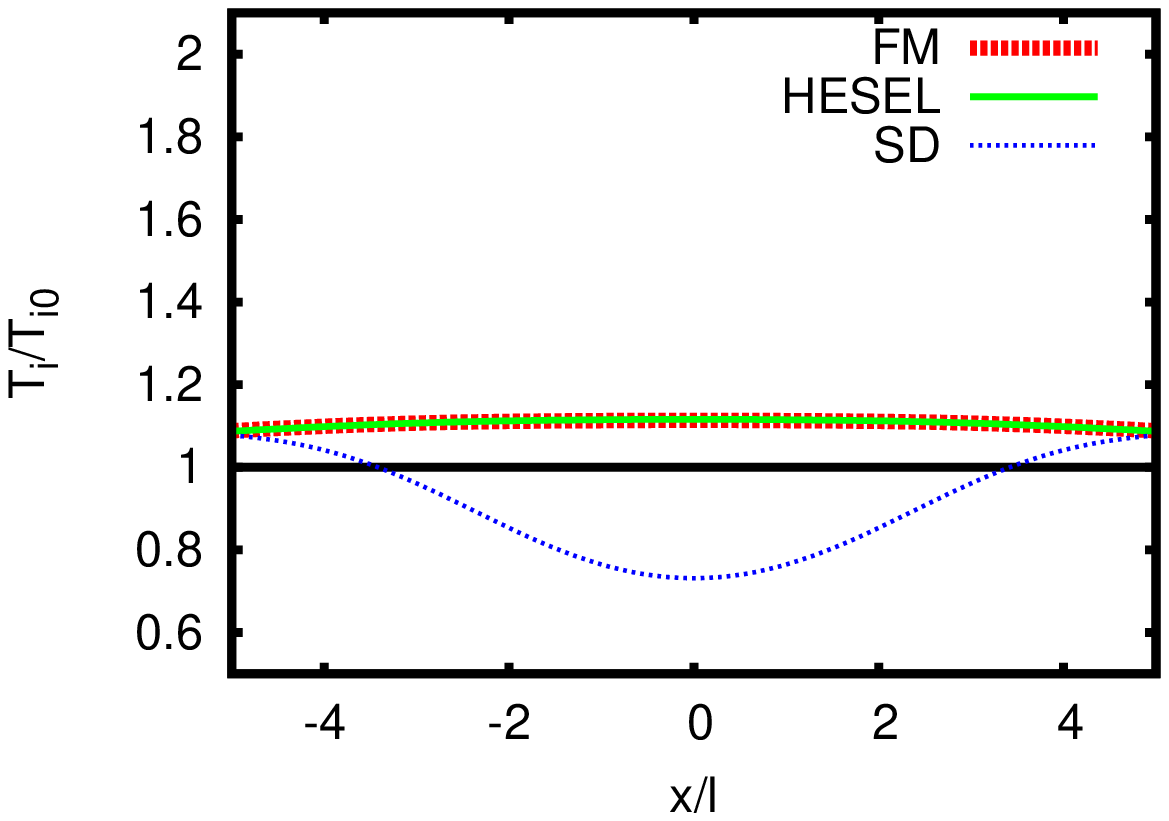}
\label{fig:ti} 
\end{minipage}
 \caption{1D simulations of perpendicular collisional diffusion (no turbulence). Radial structure of particle density, electron and ion pressures and temperatures at $t = 1.0 \unit{t_{\text{d}}}$ solving the full non-linear model (FM) Eqs.~\reff{eq:FM} (green), the HESEL model given in Eqs.~ \reff{eq:HESEL} (blue), and a simple diffusion model given in Eqs.~\reff{eq:SD} (purple). Solid black lines shows the initial condition $t=0$ given in  Eq.~\reff{eq:initialcondition}. Only one third of the simulation domain is shown.} 
\label{fig:diff_solutions}
\end{figure}

For comparison we have also solved simple diffusion equations with constant diffusion coefficients $D_e$ and $D_i$ as in the HESEL model 
\begin{align}
	\left.
	\begin{aligned}
		\pfrac{}{t}n &= D_e (1+\tau) \np^2 n, \\
		\frac{3}{2}\pfrac{}{t}p_e &= D_e(1+\tau) \np^2 p_e, \\
		\frac{3}{2}\pfrac{}{t}p_i &= 2D_i \np^2 p_i. 		
	\end{aligned}
	\right\}\quad \text{Simple Diffusion (SD)}
	\label{eq:SD}
\end{align} 
The fields are initialized as 
\begin{align}
	n(x,t=0) = n_0 (1+ e^{-\frac{x^2}{2l^2}}),\quad
	T_e(x,t=0) = T_{e0}  (1+ e^{-\frac{x^2}{2l^2}}),\quad
	T_i(x,t=0) = T_{i0}  
	\label{eq:initialcondition}
\end{align}
and the constant reference values used in the HESEL and simple diffusion model equations were 
\begin{align}
n_0 = 1.5 \times 10^{19} \unit{m^{-3}}, \quad
T_{e0} = T_{i0} = 10 \unit{eV}, \quad
B_{0} = 2 \unit{T}, \quad
m_i = 2 \unit{m_p}
\end{align}
where $m_p$ denotes the proton mass. In figure \ref{fig:diff_solutions}
we show solutions of all three set of equations at $t  = t_{\text{d}}$ for $l = 7.1 \unit{\rho_s}$. $t_{\text{d}} = \frac{3}{2}\frac{l^2}{(1+\tau)D_e} = 8.2 \times 10^5  \Omega_{ci}^{-1}$ is the time at which the particle density amplitude is half the initial value according to the analytical solution of the 1D diffusion equation.  We have investigated various initial conditions but have chosen this specific case where the initial temperatures are not identical in order to demonstrate the energy exchange mechanisms within the model. The size $L$ of the simulation domain is $L = 30 l$. In the simulations the time evolution of the particle density $n$ does not vary significantly between the three models. The SD and HESEL models give almost identical results whereas the FM model shows a slightly flatter profile. For the electron and ion temperatures and pressures the FM model and the HESEL models agree quite well. The SD model profiles, on the other hand, significantly differs from the profiles of the HESEL and FM models. The presence of energy exchange terms in the HESEL and FM models is evident. Here, the electron and ion temperatures are equilibrated whereas the absence of an energy coupling in the SD model is evident,  particularly in the ion temperature. In conclusion, the HESEL model matches the full model quite well. The simulations also demonstrate that a simple diffusion model should be used with care. The implications of the choice of perpendicular collisional model in turbulence simulations is not investigated here but is left for future work. Lastly, in the 1D simulations we have observed that solving the full model takes  significantly longer time than solving the HESEL model. A comparison between the computational requirements of the full model FM and HESEL is likewise left for future work.

\section{Conclusions}\label{sec:conclusions}
In this paper we have demonstrated how perpendicular collisional effects can be incorporated in low-frequency drift-fluid turbulence models. Based on the electrostatic limit of the Braginskii fluid equations\cite{Braginskii}, it was shown how resistivity and viscosity give rise to perpendicular fluid drifts which when inserted into the fluid moment equations provide an energy conserving closed model governing the time-evolution of the particle density, vorticity, and electron and ion pressure. We demonstrated that resistivity gives rise to energy exchange between the electron and ion thermal energies and that viscous effects due to ion-ion collisions transfer energy between the modified kinetic and the ion thermal energies. It was also shown that collisions do not imply energy sinks and sources in the global energy theorem, but merely provide conservative energy transfer terms. In conclusion, it was shown that the inclusion of perpendicular collisional transport in drift-fluid models gives rise to perpendicular diffusion in all moment equations, which obviates the need for explicit artificial diffusion. Based on the general results, we also derived a simplified 2D model aiming at describing interchange driven turbulent transport in the vicinity of the outboard midplane. The model is named HESEL and, as the predecessor ESEL, it is aimed at being computational efficient allowing the model equations to be time integrated for multiple turbulence de-correlation times. The validity of the simplified collisional dynamics in HESEL was investigated by means of numerical simulations without turbulence. The simulations showed good agreement between HESEL and the full model. 

\acknowledgments
This work has been carried out within the framework of the EUROfusion Consortium and has received funding from the Euratom research and training programme 2014-2018 under grant agreement No. 633053. The views and opinions expressed herein do not necessarily reflect those of the European Commission. 


%
\end{document}